\documentclass[aps,prb,twocolumn,superscriptaddress,showpacs,floatfix,a4paper]{revtex4}

\usepackage{amsmath}
\usepackage{amssymb}
\usepackage{epsfig}
\usepackage{multirow}
\usepackage{graphicx}

\newcommand {\dr}{{\mathrm d}\mathbf{r}}

\newcommand {\dd}{{\mathrm d}}
\newcommand {\rr}{\mathbf{r}}
\newcommand {\drr}{{\mathrm d}\mathbf{r}}
\newcommand {\kk}{\mathbf{k}}

\newcommand {\rv}{{\bf r}}

\newcommand {\jj}{\mathbf{j}}
\newcommand {\PP}{\mathbf{P}}
\newcommand {\pp}{\mathbf{p}}

\newcommand{\subrm}[1]{\ensuremath{_{\mathrm{#1}}}}

\newcommand{\medcirc}{\circ}

\begin{document}

\title{The van Hove distribution function for Brownian hard spheres:\\
dynamical test particle theory and computer simulations for bulk dynamics}

\author{Paul\ Hopkins}
\email[]{Paul.Hopkins@bristol.ac.uk}
\affiliation{H.H.\ Wills Physics Laboratory, University of Bristol, Tyndall Avenue, Bristol BS8 1TL, UK}

\author{Andrea Fortini}
\affiliation{Theoretische Physik II, Physikalisches Institut Universit{\"a}t Bayreuth, D-95440 Bayreuth, Germany}

\author{Andrew J.\ Archer}
\email[]{A.J.Archer@lboro.ac.uk}
\affiliation{Department of Mathematical Sciences, Loughborough University, Loughborough LE11 3TU, UK}

\author{Matthias Schmidt}
\affiliation{H.H.\ Wills Physics Laboratory, University of Bristol, Tyndall Avenue, Bristol BS8 1TL, UK}
\affiliation{Theoretische Physik II, Physikalisches Institut Universit{\"a}t Bayreuth, D-95440 Bayreuth, Germany}

\date{\today}


\begin{abstract}
We describe a test particle approach based on dynamical density functional theory (DDFT)
for studying the correlated time evolution of the particles that constitute a fluid. 
Our theory provides a means of calculating the van Hove
distribution function by treating its self and distinct
parts as the two components of a binary fluid mixture, with the `self'
component having only one particle, the `distinct' component consisting
of all the other particles, and using DDFT to
calculate the time evolution of the density profiles for the two components. We apply this
approach to a bulk fluid of Brownian hard spheres and compare to results for  the van
Hove function and the intermediate scattering function from Brownian dynamics computer simulations. We find good agreement
at low and intermediate densities using the
very simple Ramakrishnan-Yussouff [Phys.\ Rev.\ B {\bf 19}, 2775 (1979)]
approximation for the excess free energy functional. Since the DDFT is based on the equilibrium 
Helmholtz free energy functional, we can probe a free energy
landscape that underlies the dynamics. Within the mean-field approximation we find that as the particle density
increases, this landscape develops a minimum, while an exact treatment of a model confined situation shows that for an ergodic fluid this landscape should be monotonic. We discuss possible implications for slow, glassy and arrested dynamics at high densities.
\end{abstract}

\maketitle



\section{Introduction}

The structure of condensed matter is commonly probed by X-ray
or neutron scattering techniques, that yield quantities such as $S(k)$, the
static structure factor, and its dynamical counterpart $F(k,t)$, the intermediate
scattering function \cite{hansen2006tsl}. However, in recent years, with the
advent of modern confocal microscopes, which are able to characterise the
structure of colloidal suspensions in real-space, there is an equally great
emphasis on determining the radial distribution function $g(r)$ and its dynamical
counterpart $G(r,t)$, the van Hove distribution function
\cite{hansen2006tsl,vanhove1954csa,kegel2000dod}. 
The van Hove distribution function $G(r,t)$ is a real-space dynamical correlation
function for characterising the spatial and temporal distributions of pairs of particles
in a fluid. It gives the probability of finding a
particle at position $\rr$ at time $t$, where $|\rr|=r$, given that one of the particles was located
at the origin at time $t=0$. The intermediate scattering function $F(k,t)$ is simply obtained
from $G(r,t)$ via spatial (three-dimensional) Fourier transform. 
Pair correlation functions are important because of the significant amount of
information that they contain: Transport coefficients can be calculated
via Kubo formulae -- for example the diffusion coefficient $D$ can be obtained from $G(r,t)$ --
and thermodynamic quantities such as the
internal energy and the pressure can be related to spatial integrals\cite{hansen2006tsl} involving $g(r)$. Whether or not a liquid is near to freezing can often be discerned from
inspecting the height of the principal peak in $S(k)$: it was first noticed by
Hansen and Verlet \cite{HansenVerlet} that many simple liquids freeze when
the principal peak in $S(k)$ at $k=k_m$, takes the value $S(k_m)\simeq 2.85$. Whether a system is a glass (i.e.\ an amorphous solid) rather than a fluid may be
determined from the long time
limit value of $F(k,t)$, because in a fluid the $t \to  \infty$ limit of $F(k,t)$ is zero, whereas for
a glass this limit takes  non-zero values. This brief (and incomplete) survey is intended to
demonstrate that both dynamical and static pair correlation functions
are fundamental for characterising and understanding liquids.

 In the history of liquid state physics, fluids of hard spheres have proved to
 be an important model system for developing new techniques and theories.
 The hard sphere model is composed of particles interacting via the pair
 potential
\begin{equation} 
v_{hs}(r)=\left \{ 
\begin{array}{ll}
\infty  & \hspace{1cm}  r<  \sigma \\
0 & \hspace{1cm} {\rm otherwise},
\end{array} \right . 
\label{eq:hs_pot}
\end{equation}
where $r$ is the distance between the centers of the particles and
$\sigma$ is the hard sphere diameter. Hard spheres play an important role in describing real systems,
because attractive interactions such as those present in the Lennard-Jones potential,
can often be treated as a perturbation to the hard sphere
system \cite{hansen2006tsl}. Hence a theory that can successfully describe the properties of the
hard sphere fluid forms a good candidate to work for more realistic systems.
The hard sphere model has further grown in importance in recent decades
due to the fact that 
Eq.~\eqref{eq:hs_pot} provides a good model for the effective interaction
potential between colloidal particles in suspension, in the case when the charges
on the colloids are small or well screened -- see e.g.\ Ref.\
\onlinecite{pusey1986pbc} for an example of such a system. As the density of a
hard sphere fluid is increased, the system freezes
to a crystalline state, and the hard sphere model has played an important role in
developing our understanding of this phase transition. In contrast, although the glass transition has
attracted much interest in recent years, it is still not completely understood.
An introduction to the vast literature on this subject can be found in
Refs.\ \onlinecite{hansen2006tsl, Gotze1989} and references therein.
A number of universal processes have been discovered, including dynamical
heterogeneity \cite{weeks2002pcr}, stretched exponential decay of correlation
functions \cite{berthier2005dee}, and two-stage relaxation times \cite{vanmegen1993gtc}.
In order to understand the processes involved in structural arrest, a number
of different theoretical approaches have been used. In particular, mode-coupling theory (MCT)
has been successful in describing the bulk glass transition for hard sphere colloids
\cite{vanmegen1993gtc}, and has been applied e.g. to suspension rheology~\cite{brader2007dcs,brader2008fpc}. Nevertheless, alternatives to MCT have been developed \cite{YRMN2000pre,YRMN2001pre, MN2003pre, CRMN2006physicaA,MN2009jpcm}.
What is clear from the many studies of
arrested systems, is that  key signatures of the slow dynamics are manifest
in dynamical pair correlation functions.

In our previous Rapid Communication \cite{Archer2007dynamics},
a theory to calculate the van Hove function was proposed. For a bulk fluid of
particles interacting via Gaussian pair potentials, comparison with Brownian
dynamics (BD) computer
simulation results showed that the theory is very reliable for determining
$G(r,t)$ for this particular model system.
The theory is formulated for inhomogeneous systems and hence was also
applied to investigate the dynamics of hard spheres confined between two
parallel hard walls \cite{Archer2007dynamics}. This approach has since been
applied to investigate dynamics in liquid crystalline systems \cite{bier2008}.
In the current paper we explore the theory further, and apply it to study a bulk fluid of Brownian hard spheres.
We present results for the self and distinct parts of the
van Hove distribution function, $G_s(r,t)$ and $G_d(r,t)$ respectively, 
and by Fourier transforming, for the intermediate scattering function $F(k,t)$. We also display results for the scaled intermediate scattering function $\phi(k,t) \equiv F_s(k,t)/F_s(k,t=0)$ evaluated at the wave number $k\sigma=2\pi$. This function is often the central object of focus of MCT
\cite{Gotze1989}. The two stage relaxation of $\phi(k,t)$, that MCT predicts close to the glass transition, is also
present in our theory. 
We also determine $G(r,t)$ and $F(k,t)$ using BD computer
simulations and compare these results with those from the theory.

Our starting point is a dynamical
generalisation of Percus' test-particle approach \cite{percus1962amc}
for determining the radial distribution function $g(r)$.
Percus showed that for a fluid of classical particles interacting via
the pair interaction potential $v(r)$, that if one sets the external
potential acting on the fluid $u(\rr)=v(r)$, then the one body
density distribution $ \rho(r)$ of the fluid around the fixed `test'
particle is equal to the radial distribution function, multiplied by the
bulk fluid density $\rho $; i.e.\ Percus showed that
$\rho(r)=\rho  g(r)$. When using equilibrium density functional
theory (DFT) \cite{evans1992fif, hansen2006tsl} to study a fluid,
the test-particle method provides a useful route to obtain $g(r)$ because $u(\rr)$ (and hence $v(r)$) enters the framework explicitly. We also describe an alternative `zero-dimensionality' approach for calculating $g(r)$. This forms a stepping stone in the development of the dynamical theory.

We apply a dynamical extension of Percus' idea, together with dynamical
density functional theory (DDFT)
\cite{marconi1999uat, marconi2000ddf, archer2004ddf} in order
to calculate the van Hove function $G(r,t)$ in general inhomogeneous
situations. We implement the method using the very
simple Ramakrishnan-Youssouff (RY) approximation for the Helmholtz
free energy functional \cite{ramakrishnan1979fpo}. We find
that the results from the theory agree well with those from BD
computer simulations when the fluid density
$\rho  \sigma^3 \lesssim 0.6$. At higher densities the free energy underlying the dynamics
develops a minimum, corresponding to the appearance of a free
energy barrier that must be traversed for a particle to escape from
the cage formed by the neighbouring particles.

In addition, we compare our results for $G(r,t)$ to those obtained by assuming that
$G_s(r,t)$ takes a simple Gaussian form for all times $t$, together with the
Vineyard approximation \cite{hansen2006tsl, vineyard1958scattering},
which sets $G_d(r,t)$
to be a simple convolution of $G_s(r,t)$ and the radial distribution
function $g(r)$, as described in detail below.
We find that in contrast to the received wisdom \cite{Hansen1986}, the simple
Vineyard approximation is actually a fairly good approximation for
the van Hove function for Brownian hard sphere fluids at low and intermediate
densities.

We also compare to an equilibrium DFT based approach with which we are
able to calculate a series of density profiles that agree well with
those from the DDFT. This is done by performing a constrained
minimisation of the free energy through the judicious choice of
a suitable external potential to confine the test particle. This
approach is easier to implement than the full DDFT and allows
for the free energy landscape underlying the dynamics to be
mapped out and examined in detail. However, this approach
does not give any of the time information that the full DDFT
gives -- i.e.\ it yields the van Hove function with the time labels
`removed'. One of the advantages of this approach is that for a particular
(parabolic) choice of external potentials, we are able
to calculate exactly the fluid density profiles, which are precisely those
predicted by Vineyard's theory \cite{hansen2006tsl, vineyard1958scattering}.
We discuss the significance of this result below, after we have
laid out the general structure of the theory and shown the results.

This paper is structured as follows: In Sec.\ \ref{sec:back} we outline the necessary theoretical background, including the definition of the van Hove function, the Vineyard approximation, DDFT, and the static test particle limit. Most of this section may be safely skipped by expert readers. In Sec.~\ref{sec:test_part_lim} the dynamical test particle limit is introduced. Sec.~\ref{sec:hard_spheres} summarises the model used and describes the simulation details. In Sec.~\ref{sec:results} we describe results from the different dynamical approaches, the corresponding equilibrium approaches, and the free energy landscape. In Sec.~\ref{sec:conclusion} we make some concluding remarks. Appendix \ref{app:A} presents an exact solution of a corresponding equilibrium situation.

\section{Background}
\label{sec:back}

\subsection{The van Hove function}
\label{sec:vanhove}
We first recall the definition of the van Hove function and some
of its properties; for a more detailed account see Refs.\
\onlinecite{hansen2006tsl,vanhove1954csa}.  Consider a set of $N$
particles with time dependent position coordinates $\rr_i(t)$,
where $i=1,..,N$ is the particle index, and $t$ is time. The van
Hove correlation function is defined as the probability of finding
a particle at position $\rr$ at time $t$, given that there was a particle
at the origin at time $t=0$: 
\begin{equation}
 G(r,t)=\frac{1}{N}\left\langle\sum_{i=1}^N\sum_{j=1}^N\delta(\rr+\rr_j(0)-\rr_i(t))\right\rangle,
\label{eq:van_hove_2}
\end{equation}
where $\langle\cdot\rangle$ represents an ensemble average
and $\delta(\cdot)$ is the  three-dimensional Dirac delta function. $G(r,t)$ can be
naturally separated into two terms, conventionally referred to
as its ``self'' and ``distinct'' part, by discriminating between the
cases $i=j$ and $i\neq j$, respectively. So
\begin{eqnarray}
 G(r,t)&=&\frac{1}{N}\left\langle\sum_{i=1}^N\delta(\rr+\rr_i(0)-\rr_i(t))\right\rangle \nonumber \\
 &&+\frac{1}{N}\left\langle\sum_{i\neq j}^N\delta(\rr+\rr_j(0)-\rr_i(t))\right\rangle \nonumber \\
 &\equiv&G_s(r,t)+G_d(r,t),
\label{eq:van_hove_3}
\end{eqnarray}
where the self part, $G_s(r,t)$, describes the average motion of the
particle that was initially at the origin, whereas the distinct part,
$G_d(r,t)$, describes the behavior of the remaining $N-1$
particles. At $t=0$, Eq.~(\ref{eq:van_hove_3}) reduces to the static
particle-particle auto-correlation function, which is defined as
\begin{eqnarray}
G(\rr,0)&=&\delta(\rr)+\frac{1}{N}\left\langle \sum_{i\neq j}^N\delta(\rr+\rr_j(0)-\rr_i(0))\right\rangle \nonumber \\
&= & \delta(\rr)+\rho g(\rr),
\label{eq:van_hove_3.1}
\end{eqnarray}
where $g(\rr)$ is the (static) pair distribution function. For the
homogeneous bulk fluid $\rho(\rr)=\rho $; isotropy implies that
the dependence is only on $r=|\rr|$. Thus, at $t=0$:
\begin{eqnarray}
  G_s(r,0)&=& \delta(\rr)  \label{eq:van_hove_3.5}\\
  G_d(r,0)&=&\rho  g(r).
\label{eq:van_hove_4}
\end{eqnarray}
From the definitions of $G_s(r,t)$ and $G_d(r,t)$,
Eq.\ (\ref{eq:van_hove_3}), it is clear that the volume integral of
these functions must be a conserved quantity for all times $t$:
\begin{eqnarray}
 \int \dr G_s(r,t)&=&1, \label{eq:van_hove_4.5}\\
 \int \dr G_d(r,t)&=&N-1. 
\label{eq:van_hove_5}
\end{eqnarray}
The asymptotic behaviour of $G(r,t)$ in bulk in the thermodynamic limit
is obtained by considering $N\to\infty$ and volume $V\to\infty$ such
that $N/V=\rho $ is finite:
\begin{align}
\lim_{r\to\infty}G_s(r,t)=\lim_{t\to\infty}G_s(r,t)=0,\label{eq:van_hove_7.5} \\ 
\lim_{r\to\infty}G_d(r,t)=\lim_{t\to\infty}G_d(r,t)=\rho .  
\label{eq:van_hove_7}
\end{align}
A key quantity that we use below to characterise $G_s(r,t)$ is its width $w(t)$ defined via
\begin{equation}
(w(t))^2 = 4\pi\int_0^\infty \dd r \; r^4 G_s(r,t) ,
\label{eq:width}
\end{equation}
the second moment of $G_s(r,t)$.
It is often convenient to consider the intermediate scattering function
which is related to the van Hove function via a spatial Fourier transform,
\begin{equation}
F(k,t) = \int \drr \,G(r,t) \exp(-i \kk \cdot \rr).
\label{eq:FT}
\end{equation}
This quantity is directly accessible in light and neutron scattering experiments
\cite{hansen2006tsl}. 

\subsection{Approximating $G_s(r,t)$}

A commonly used approximation for the self part of the van Hove function is
to assume a Gaussian shape \cite{hansen2006tsl}:
\begin{equation}
G_s(r,t)=\frac{1}{\pi^{3/2}W(t)^3}\exp\left( -\frac{r^2}{W(t)^2}\right)
\label{eq:Gs_gauss}
\end{equation}
where the width $W(t)=\sqrt{\frac{3}{2}} w(t)$, when $w(t)$ is calculated via
Eq.\ \eqref{eq:width}. The form \eqref{eq:Gs_gauss} is exact in the limits
$t \to 0$ and $t \to \infty$ for all densities when the system is fluid \cite{hansen2006tsl}. 
It is also
exact for all times $t$ in the low density limit $\rho  \to 0$, where
interactions between the particles can be neglected. There are a
number of approximations for $W(t)$. In molecular dynamics, at very short
times $t\ll\tau_c$, where $\tau_c$ is the mean collision time, particles in a
fluid do not experience collisions and therefore move freely at a constant
velocity. This is akin to an ideal gas where the particle velocities follow
a simple Maxwellian (Gaussian) distribution, giving 
\begin{equation}
W(t) = t\sqrt{2/\beta m},
\label{eq:w_gas}
\end{equation}
where $m$ is the particle mass. Over longer times $t \gg \tau_c$ the
particles in the fluid undergo many collisions with neighbouring particles,
so that the trajectory of a given particle is a random walk and thus its
probability distribution $G_s(r,t)$ is the solution of the diffusion equation
\begin{equation}
\frac{\partial G_s(\rr,t)}{\partial t}=D_l\nabla^2G_s(\rr,t),
\label{eq:diffusion}
\end{equation}
where $D_l$ is the long time self diffusion coefficient. For the Dirac delta
initial condition \eqref{eq:van_hove_3.5}, the solution of Eq.~\eqref{eq:diffusion}
corresponds to the Gaussian form \eqref{eq:Gs_gauss}, with 
\begin{equation}
W(t) = 2\sqrt{D_lt}.
\label{eq:diff_atom}
\end{equation}

For colloidal particles, the collisions with the solvent atoms happen so frequently that the time scale $\tau_c$ is much smaller than all other time scales relevant for the dynamics, such as the Brownian time scale $\tau_B$ which is roughly the time for a particle to travel a distance equal to its own diameter, and also the typical time scale between collisions of pairs of colloids, $\tau_{col}$. This means that we may set $\tau_c \to 0$ and that for a low density suspension of colloids Eq.\ \eqref{eq:diff_atom} holds for all times $t$. Thus, we may combine Eqs.\ \eqref{eq:Gs_gauss} and \eqref{eq:diff_atom} to obtain
\begin{equation}
G_s(r,t)=(4\pi D_lt)^{-3/2}\exp\left( -\frac{r^2}{4D_lt}\right).
\label{eq:Gs_gauss_colloid}
\end{equation}
We find below that for Brownian hard spheres this approximation is not only reliable in the low density limit, but is also fairly good up to intermediate densities $\rho \sigma^3 \lesssim 0.6$.

\subsection{Vineyard approximation for $G_d(r,t)$}
\label{sec:vineyard_appr}
Vineyard \cite{vineyard1958scattering}
suggested that one may rewrite the distinct part of the van Hove function as
\begin{equation}
G_d(r,t) = \int\dr'g(\rr')H(\rr,\rr',t),
\label{eq:convol}
\end{equation}
which is merely a redefinition of $G_d(r,t)$ in terms of the unknown function $H(\rr,\rr',t)$, which is the probability
that if there was a particle at the origin at time $t=0$ and a second particle located at $\rr'$,
this second particle is later located at $\rr$ at time $t$. Vineyard's approximation is to
replace $H(\rr,\rr',t)$ by $G_s(\rr-\rr',t)$, giving
\begin{equation}
G_d(r,t) = \int\dr'g(\rr')G_s(\rr-\rr',t).
\label{eq:vineyard_convol}
\end{equation}
Some comments in the literature state that the Vineyard approximation ignores important
correlations that inhibit the rate at which the structure of the liquid breaks up
and it therefore predicts too rapid decay of this structure \cite{Hansen1986}. This may indeed be the case for fluids undergoing molecular dynamics,
but for the system with Brownian dynamics (over-damped stochastic equations of motion)
that we consider here, we find that
taking Eq.\ \eqref{eq:Gs_gauss_colloid} together with Eq.\ \eqref{eq:vineyard_convol},
is actually fairly reliable -- in particular when the fluid is at low and
intermediate densities $\rho  \sigma^3 \lesssim 0.6$. We will henceforth
refer to Eqs.\ \eqref{eq:Gs_gauss_colloid} and \eqref{eq:vineyard_convol}
as the `Vineyard approximation' for the van Hove function.

\subsection{DDFT and equilibrium DFT}
\label{sec:DDFT_equilibrium DFT}

The dynamics of a system of $N$ Brownian (colloidal) particles with
positions $\rr_i(t)$ can be modelled with the following set of
(over-damped) stochastic equations of motion \cite{dhont1996idc}:
\begin{equation}
\Gamma^{-1}\frac{\dd \rr_i(t)}{\dd t}=-\nabla_i U_N(\rr^N,t)+ \zeta_i(t),
\label{eq:EOM}
\end{equation}
where $\rr^N=\{\rr_i;\, i=1,\ldots,N \}$ is the set of particle
coordinates, $\Gamma^{-1}$ is a friction constant characterizing the
one-body drag of the solvent on the particles, $\zeta_i(t)$ is a
stochastic white noise term and the total inter-particle potential
energy is
\begin{eqnarray}
U_N(\rr^N,t)
=\sum_{i=1}^N u(\rr_i,t) 
+ \frac{1}{2} \sum_{i=1}^N  \sum_{j \neq i} v(|\rr_i-\rr_j|),
\end{eqnarray}
which is composed of a one-body contribution due to the external
potential $u$ (which may or may not be time dependent), and a
sum of contributions from the pair interactions between the particles. The time
evolution of the probability density for the particle coordinates
$P^{(N)}(\rr^N,t)$ is described by the Smoluchowski equation
\cite{dhont1996idc,archer2004ddf}:
\begin{eqnarray}
  \frac{\partial P^{(N)}}{\partial t} = \Gamma \sum_{i=1}^N
  \nabla_i \cdot [ k_BT \nabla_i P^{(N)}+ \nabla_i U_N P^{(N)} ].
  \label{eq:smoluchowski}
\end{eqnarray}
The one-body density is obtained by integrating over the position
coordinates of all but one particle:
\begin{equation}
\rho(\rr_1,t) \, =\, N \int \dr_2 \, ... \int \dr_N P(\rr^N,t).
\label{eq:intrho}
\end{equation}

Integrating the Smoluchowski equation (\ref{eq:smoluchowski}) we
obtain the key equation of DDFT \cite{archer2004ddf}:
\begin{align}
  \frac{\partial \rho(\rr,t)}{\partial t} &= \Gamma \nabla \cdot
  \left[\rho(\rr,t) \nabla \frac{\delta F[\rho(\rr,t)]}{\delta
      \rho(\rr,t)}\right],
\label{eq:DDFT}
\end{align}
where $F[\rho]$ is taken to be the {\em equilibrium} total Helmholtz free
energy functional:
\begin{eqnarray}
  F[\rho(\rr)] = k_BT \int \dr \rho(\rr)
  [\ln(\Lambda^3\rho(\rr))-1] \notag \\
  +F\subrm{ex}[\rho(\rr)]
  +\int \dr \, u(\rr) \rho(\rr),
\label{eq:F_one}
\end{eqnarray}
where the first term on the right hand side is the ideal-gas contribution
to the free energy, $\Lambda$ is the thermal de Broglie wavelength,
$F\subrm{ex}[\rho(\rr)]$ is the excess (over ideal gas)
part of the free energy, which is in general unknown exactly, and we
have suppressed the dependence on temperature $T$ and volume $V$ in
the notation. In obtaining (\ref{eq:DDFT}) we have made the
approximation that equal-time two-body correlations at each time $t$ in the
non-equilibrium situation are the same as those of an equilibrium fluid
with the same one-body density profile $\rho(\rr,t)$, generated by an
appropriate external potential \cite{marconi1999uat, marconi2000ddf,
  archer2004ddf}. It has been shown in a variety of cases that the
DDFT (\ref{eq:DDFT}) is reliable in predicting the time-evolution of
$\rho(\rr,t)$, when solved in conjunction with a sufficiently accurate approximation
for the equilibrium Helmholtz free energy functional $F[\rho(\rr)]$ --
see for example the results presented in Refs.\ \onlinecite{marconi1999uat,
  dzubiella2003mfd, penna2003ddf, archer2005ddf, archer2005ddf_b,
  rex2005scd, rex2006ucc, royall2007nsc, rex2007ddf, rauscher2007ddf}.

Although in the following we will not go beyond dynamics that are local
in time, it is worth mentioning that 
more generally, going beyond the case of particles with stochastic
over-damped equations of motion (\ref{eq:EOM}), Chan and Finken
\cite{chan2005tdd} established a rigorous DDFT for classical fluids,
showing that the time evolution of the one body density $\rho(\rr,t)$
is obtained from the solution of
\begin{eqnarray}
\frac{\partial \rho(\rr,t)}{\partial t}=-\nabla \cdot \jj(\rr,t),
\label{eq:continuity} \\
\frac{\partial \jj (\rr,t)}{\partial t}=\PP[\rho(\rr,t)],
\label{eq:current_eq}
\end{eqnarray}
where $\jj(r,t)$ is the particle current density, and
Eq.\ (\ref{eq:continuity}) represents a continuity equation for the
one-body density $\rho(\rr,t)$. One should, of course, expect on
general grounds for the dynamical equations to be of this
{\em form}\cite{Kreuzer, ChaikinLubensky,Russeletal} -- recall
that Eq.\ (\ref{eq:current_eq}) is the continuity equation. However,
the functional $\PP[\rho(\rr,t)]$ that governs the
time evolution of $\rho(\rr,t)$, takes a form that depends on the equations
of motion of the particles -- i.e.\ it depends on whether the
particles evolve under Newtonian dynamics or have stochastic equations
of motion such as (\ref{eq:EOM}).

Due to the fact that in general the
functional $\PP[\rho(\rr,t)]$ in Eq.\ (\ref{eq:current_eq}) is an unknown
quantity, one is prevented from applying our DDFT approach for
calculating dynamic correlation functions, and we are restricted to
the the Brownian case (\ref{eq:EOM}) outlined above. The particular
approximation used in Eqs.\ (\ref{eq:continuity}) and
(\ref{eq:current_eq}) to obtain Eq.\ (\ref{eq:DDFT}), is to assume the
one particle current density to be of the form
\begin{equation}
\jj(\rr,t)=-\Gamma \rho(\rr,t) \nabla \frac{\delta F[\rho(\rr,t)]}{\delta\rho(\rr,t)}.
\label{eq:current}
\end{equation}
Nevertheless, there is much active research aimed at going beyond the
simple overdamped case \cite{marconi2006nid, archer2006ddf, marconi2007psa,
Marconietal2008jpcm, Marconietal2009jcp, archer2009jcp}.

In what follows we will relate the van Hove function to the time
evolution of the one-body density profiles of a binary mixture; we
therefore require the multicomponent generalization\cite{archer2005ddf} of Eq.\ (\ref{eq:DDFT}):
\begin{align}
\frac{\partial \rho_i(\rr,t)}{\partial t} &= \Gamma \nabla \cdot
\left[\rho_i(\rr,t) \nabla \frac{\delta F[\{\rho_i\}]}{\delta
\rho_i(\rr,t)}\right],
\label{eq:MT_DDFT}
\end{align}
where $F[\{\rho_i\}]$ has the following form
(cf.\ Eq.~\eqref{eq:F_one}) \cite{footnote1}:
\begin{align}
F[\{\rho_i\}]=k_BT \sum_{i} \int \dr\, \rho_i(\rv)[\ln \Lambda^3\rho_i(\rv)-1]\notag \\
+F_{\mathrm ex}[\{\rho_i\}] +\sum_{i}\int \dr \, u_i(\rv) \rho_i(\rv).
\label{eq:F}
\end{align}
where the summations run over all species $i$.  Given an initial set
of density profiles, $\{\rho_i(\rr,t=0)\}$, we may employ the DDFT
equations (\ref{eq:MT_DDFT}) and (\ref{eq:F}) to calculate the full
time evolution of the one-body density profiles $\rho_i(\rr,t)$.

For completeness, we recall some of the key results from
equilibrium  \cite{evans1992fif, hansen2006tsl}.  For a given set of
(one-body) external potentials
$\{u_{i}(\rr)\}$, the unique set of one-body density profiles
$\{\rho_i(\rr)\}$ are those which minimize the Helmholtz free energy
of the system $F[\{\rho_i\}]$, subject to the constraint that the
average number of particles of each species $\int \dr
\rho_i(\rr)=N_i$, is fixed. This is equivalent to an unconstrained
minimization of the grand potential functional
\begin{equation}
  \Omega[\{\rho_i\}]=F[\{\rho_i\}]-
  \sum_{i}\mu_i \int \dr \rho_i(\rr),
\label{eq:Omega}
\end{equation}
where the Lagrange multiplier $\mu_i$ is the chemical potential of
species $i$. 
Minimization with respect to variations in the density profiles yields
the following set of Euler-Lagrange equations
\cite{hansen2006tsl,evans1992fif}:
\begin{equation}
\frac{\delta F[\{\rho_i\}]}{\delta \rho_i(r)} = \mu_i.
\label{eq:F_min}
\end{equation}
The Euler-Lagrange equations can be rewritten as
\begin{equation}
 \rho_i(\rr)=\Lambda^{-3} \exp 
 \bigg[ \beta \mu_i-\beta u_i(\rr)+c_i^{(1)}(\rr[\{\rho_j\}]) \bigg],
\label{eq:app_1}
\end{equation}
where
\begin{equation}
  c_i^{(1)} (\rr; [\{\rho_j\}])=
  -\beta \frac{\delta F\subrm{ex}[\{\rho_j\}]}{\delta \rho_i(\rr)},
\label{eq:c_1}
\end{equation}
is the one-body direct correlation functional. The set of density
profiles that satisfy \eqref{eq:app_1} minimize the free energy and
are the equilibrium density profiles. When the equilibrium set
of profiles $\{ \rho_i(\rr) \}$ are substituted into \eqref{eq:Omega}, 
 the grand potential $\Omega$ of the system is obtained. 

\subsection{Percus' test particle limit}
\label{sec:def_test_part_lim_details}

Here we give a derivation of Percus' (static) test particle limit
closely following Ref.~\onlinecite{oettel2005ies}. Consider a one component
system such that the Helmholtz free energy, $F$, is given by
\eqref{eq:F_one}. We are interested in the change in $\rho(\rr)$ when the
external potential is changed from the potential $u'(\rr)$ to the
potential $u(\rr)$. To this end we perform a functional Taylor
expansion of $F\subrm{ex}[\rho]$ in powers of
$\Delta\rho(\rr)=\rho(\rr)-\rho'(\rr)$. For the sake of simplicity we
consider the change in going from $u'(\rr)=0$ to a
spherically symmetric external potential $u(r)$. The
variable in the Taylor expansion is then
$\Delta\rho(r)=\rho(r)-\rho^b $, where $\rho^b$ is the bulk density and
the expansion of $F\subrm{ex}[\rho]$ to second order in
$\Delta\rho(r)$ is
\begin{eqnarray}
F\subrm{ex}[\rho] &=& F\subrm{ex}[\rho^b ]+\int\dr\left.\frac{\delta F\subrm{ex}[\rho]}{\delta \rho(r)}\right\vert_{\rho^b }\Delta\rho(r) \nonumber\\
& & +\frac{1}{2}\int\dr\int\dr'\left.\frac{\delta^2 F\subrm{ex}[\rho]}{\delta \rho(r)\delta \rho(r')}\right\vert_{\rho^b }\Delta\rho(r)\Delta\rho(r') \nonumber\\ & & +O(\Delta\rho^3).
\label{eq:Fex_expans}
\end{eqnarray}
Although the form of $F\subrm{ex}$ is not specified, the functional
derivatives are related to identifiable properties of the system, so that
evaluating them at $\rho(r)=\rho^b $ gives
\begin{eqnarray}
\left.\frac{\delta F\subrm{ex}[\rho]}{\delta \rho(r)}\right\vert_{\rho^b }
&=& -k_BT \mu\subrm{ex}, \nonumber \\
\left.\frac{\delta^2 F\subrm{ex}[\rho]}{\delta \rho(r)\delta \rho(r')}\right\vert_{\rho^b }
&= &-k_BTc^{(2)}(|\rr-\rr'|), \label{eq:Fex_diff} \\
\frac{\delta O(\Delta\rho^3)}{\delta\rho(r)}&=&B(r), \nonumber
\end{eqnarray}
where $\mu\subrm{ex}$ is the excess chemical potential, $c^{(2)}(r)$
is the (pair) direct correlation function, and $B(r)$ is an unknown function
that contains the higher order terms of the Taylor expansion.  Substituting
Eqs.~\eqref{eq:Fex_expans} and \eqref{eq:Fex_diff} into \eqref{eq:F_one},
and then minimizing the functional with respect to variations in $\rho(r)$,
we obtain the following Euler-Lagrange equation [c.f.\ Eq.~\eqref{eq:F_min}]:
\begin{equation}
\frac{\delta F}{\delta \rho(r)}=\mu=k_BT\ln (\Lambda^3\rho^b )+\mu\subrm{ex},
\label{eq:F_one_min}
\end{equation}
where we have separated the chemical potential $\mu$ into an
ideal-gas and an excess (over ideal-gas) contribution $\mu\subrm{ex}$. In the case of
a spherically symmetric external potential, Eq.~\eqref{eq:F_one_min}
may be rewritten as:
\begin{eqnarray}
\frac{\rho(r)}{\rho^b }&=&\exp\Big[-\beta u(r)+\int\dr'c(|r-r'|)\Delta\rho(r') \nonumber \\
& & +B(r)\Big].
\label{eq:dft_gr}
\end{eqnarray}
For the same one-component system the bulk Ornstein-Zernike (OZ) equation
for the total correlation function, $h(r)=g(r)-1$, reads as follows:
\begin{equation}
h(r)=c(r)+\rho^b \int\dr'h(r')c(|r-r'|).
\label{eq:oz_eq}
\end{equation}
It can be shown through diagrammatic-methods \cite{hansen2006tsl} that
the OZ equation has the general solution
\begin{equation}
h(r)=c(r)+\ln(g(r))+\beta v(r)+b(r)
\label{eq:oz_sol}
\end{equation}
where $v(r)$ is the inter-particle pair potential, and $b(r)$ is the
bridge function composed of the sum of all the so-called `bridge'
diagrams \cite{hansen2006tsl}. Substituting \eqref{eq:oz_sol} into
\eqref{eq:oz_eq} we obtain:
\begin{equation}
g(r)=\exp\Big[-\beta v(r)+\int\dr'c(|r-r'|)\rho^b h(r')+b(r)\Big].
\label{eq:oz_gr}
\end{equation}
If we compare Eqs.~\eqref{eq:dft_gr} and \eqref{eq:oz_gr} we find they
have the same structure, and that they may be formally identified. If
we set $u(r)= v(r)$ in Eq.~\eqref{eq:dft_gr}, it can be
shown through diagrammatic methods \cite{hansen2006tsl, attard2002tas}
that $b(r)=B(r)$ and that
\begin{equation}
  g(r)=\rho(r)/\rho^b ,
\label{eq:percus}
\end{equation}
or alternatively, $\rho  h(r)=\Delta\rho(r)$. Thus when
$u(r)= v(r)$, Eqs.~\eqref{eq:dft_gr} and
\eqref{eq:oz_gr} become identical. Therefore we note that not
only can the OZ relationship be derived from the free energy
functional \cite{evans1992fif}, but that the equilibrium one-body
density profile in the presence of an external potential
$u(r)= v(r)$ is related to the (two-body) radial
distribution function via Eq.~\eqref{eq:percus}. We should recall at
this point that although many formal statements can be made about the
bridge function $b(r)$, in practice it is an unknown function, and all
theories for $g(r)$ constitute some form of approximation for $b(r)$
\cite{hansen2006tsl}. For example, if we set $b(r)=B(r)=0$ then
\eqref{eq:dft_gr} is equivalent to using the hyper-netted chain (HNC)
approximation \cite{hansen2006tsl} to the OZ equation. Furthermore,
Percus \cite{percus1962amc} showed that by Taylor expanding with
different functions of $\rho(r)$ one may retrieve the Percus-Yevick
and other closures to the OZ equations. This result may also be
generalized to inhomogeneous systems~\cite{oettel2005ies}.

\subsection{Zero-dimensionality route to $g(r)$}
\label{sec:zero-d}

We present an alternative method for calculating $g(r)$, although its basis is the same key idea that underpins Percus' test particle limit described above: that $g(r)$ can be obtained from the density profile of a fluid around a fixed test particle. The key difference is that instead of treating the test particle as a fixed external potential, in the zero-dimensionality route we treat the test particle via its density distribution. The density profile of a particle fixed at a point (i.e.\ in zero-dimensional space -- hence our choice of name for this limit) takes the form of a Dirac delta function. Having fixed this contribution to the density distribution [c.f.\ Eq.\ \eqref{eq:van_hove_3.1}], one can then calculate the density distribution of the remaining particles in the presence of the test particle. Specifically, we can write the grand potential functional as:
\begin{equation}
\Omega^*[\rho g(r)] = F\subrm{id}[\rho g(r)] + F\subrm{ex}[\delta(\rr)+\rho g(r))]-\mu\int\dr\rho g(r),
\label{eq:omega_star}
\end{equation}
where $\rho g(r)$ is the density distribution of the remaining particles -- the quantity we wish to calculate. Note that here, and in what follows, $\rho$ is the bulk density. The ideal gas term $F\subrm{id}$ does not contain the Dirac delta contribution -- we have crossed over to a system with $N-1$ particles. Since the bulk fluid density $\rho $ is necessarily fixed, we must simply minimise $\Omega^*$ with respect to variations in $g(r)$, giving the following Euler-Lagrange equation to be solved for $g(r)$:
\begin{equation}
\frac{\delta\Omega^*}{\delta g(r)}=0.
\end{equation}
An alternative means of calculating $g(r)$ is to treat the system as a binary mixture. The test particle (which we label `$s$'), with density distribution $\rho_s(r)=\delta(\rr)$, is one species and then we use the DFT for a binary mixture to calculate the density profile of the remain particles $\rho_d(r)$ in the presence of the density profile $\rho_s(r)$ for the fixed particle, treating the remaining particles as a second species `$d$' in the mixture. $\rho_d(r)$ is the solution of
\begin{equation}
\frac{\delta\Omega^\dag}{\delta \rho_d(r)}=0,
\end{equation}
where $\Omega^\dag$ is a modified version of Eq.\ \eqref{eq:Omega} where $\rho_s(r)=\delta(\rr)$ is fixed and therefore the ideal Helmholtz free energy, $F\subrm{id}[\rho_d]$, does not depend on $\rho_s(\rr)$, c.f. Eq.~\eqref{eq:omega_star}.


When using an approximate free energy functional, there is a difference between the the zero-dimensional limit and Percus' limit for calculating $g(r)$. This is because in the zero-dimensionality limit, in contrast to Percus' method, the test particle at the origin does not interact with the fluid via an external potential $u(r)=v(r)$, that is identical to the pair potential, but rather via an approximation $u^*(r)$ to the pair potential, generated by the approximate density functional. We calculate below in Sec.\ \ref{sec:static_structure} an explicit expression for $u^*(r)$ in the case of hard-spheres treated using the RY approximation for the free energy. We will also discuss further the relation between Percus' test particle limit and the zero-dimensionality limit.

\section{Dynamic Test Particle Limit}
\label{sec:test_part_lim}
\subsection{Definition}

We next extend the static test particle limit 
and consider the dynamical situation which allows us to calculate the van Hove function
$G(r,t)$. The key is the following observation: Consider a fixed
test particle of species `$s$' (self) located at the origin; due to Percus we know that in this situation
the density distribution of the remaining particles $\rho(r)=\rho  g(r)$.
Now consider releasing the test particle at time $t=0$ and allowing it to move through the fluid. When this happens its probability (density)
distribution $\rho_s(r,t)$ changes from a Dirac delta function (at $t=0$)
to a distribution with a non-zero value for some points away from the origin.
If we now recall the definition of the function $G_s(r,t)$ in Eq.\ \eqref{eq:van_hove_3},
we see that $G_s(r,t)$ gives the probability that a particle initially located at the origin
has moved a distance $r$ away from the origin after time $t$. Therefore,
in this situation, $\rho_s(r,t) \equiv G_s(r,t)$ for all times $t \geq 0$.
Similarly, if we consider how the remaining particles redistribute
themselves as the test particle moves away from the origin, we see from
Eq.\ \eqref{eq:van_hove_3} that the
probability of finding any one of these particles a distance $r$ from
the origin at time $t$ is given by $G_d(r,t)$. We label the
remaining particles as being particles of species `$d$' (distinct) and
having the density profile $\rho_d(r,t)$.
Thus, as in the static test particle
case, we may connect the two parts of the van Hove function with the
density profiles of a two-component system of species $s$ and
$d$:
\begin{eqnarray}
 G_s(r,t)&\equiv&\rho_s(r,t), \nonumber \\
 G_d(r,t)&\equiv&\rho_d(r,t),
\label{eq:G_rho_equiv}
\end{eqnarray}
where species $s$ is composed of only one particle, the test particle, and
$\int\dr\rho_s(r,t)=1$, so that Eq.~\eqref{eq:van_hove_4.5} is satisfied.
We may therefore formally set the pair potential for interactions between species
$s$ particles $v_{ss}(r)=0$. The density profile for species
$d$ must satisfy the normalization constraint \eqref{eq:van_hove_5},
and the self-distinct and distinct-distinct pair potentials must be
identical, $v_{sd}(r)=v_{dd}(r)=v(r)$. This is equivalent to modelling a
one-component system, but treating one particle separately from the rest.
Recall that at time $t=0$ we know the test particle's position exactly from
Eq.~\eqref{eq:van_hove_3.5} and combining this with
Eqs.~\eqref{eq:van_hove_4},  \eqref{eq:percus} and \eqref{eq:G_rho_equiv}
we obtain
\begin{eqnarray}
 G_s(r,t=0)\equiv&\rho_s(r,t=0)=&\delta(r), \nonumber \\
 G_d(r,t=0)\equiv&\rho_d(r,t=0)=&\rho  g(r).
\label{eq:G_rho_equiv2}
\end{eqnarray}

The connections made in Eq.\ \eqref{eq:G_rho_equiv} between the
self and distinct parts of the van Hove function and the density profiles
$\rho_s(r,t)$ and $\rho_d(r,t)$ in the dynamical test particle limit described
above are conceptually important. However, we have merely shifted the
problem of how to determine $G(r,t)$ onto the problem of
how to determine the time evolution of the
two coupled density profiles $\rho_s(r,t)$ and $\rho_d(r,t)$.
The solution that we use in this paper is to use DDFT, i.e. equations~\eqref{eq:MT_DDFT} are integrated forward in
time with Eqs.\ \eqref{eq:G_rho_equiv2} providing the initial time, $t=0$,
density profiles. The resulting time series of density profiles gives the
self and distinct parts of the van Hove function through Eq.\ \eqref{eq:G_rho_equiv}.
Henceforth, we refer to this as the `dynamical test particle' theory.

\subsection{Approximate solution}

Before proceeding to the results that we obtain from
following the calculation scheme described above, it is worth examining
an approximate analytical solution that may be obtained as follows:
From Eqs.\ \eqref{eq:MT_DDFT}, \eqref{eq:F} and \eqref{eq:c_1}
we may write the DDFT equations for the two density profiles
$\rho_s(r,t)$ and $\rho_d(r,t)$ as:
\begin{equation}
\frac{\partial \rho_i(\rr,t)}{\partial t} = D \nabla^2 \rho_i(\rr,t)+
\Gamma \nabla \cdot
\left[\rho_i(\rr,t) \nabla c_i^{(1)}(\rr,t) \right],
\label{eq:DDFT_TP}
\end{equation}
where $i=s,d$ and the diffusion coefficient $D=k_BT\Gamma$. If we
set the second term on the right hand side to zero and we set $D=D_l$
then we obtain Eq.\ \eqref{eq:diffusion} for $\rho_s(r,t)=G_s(r,t)$
and thus the solution to the DDFT for the self part of the van Hove function
in this limit is the Gaussian form in Eq.\ \eqref{eq:Gs_gauss_colloid}.
Similarly, for species $d$, when we assume that the second term on the right
hand side of Eq.\ \eqref{eq:DDFT_TP} can be neglected, then we obtain:
\begin{equation}
\frac{\partial \rho_d(\rr,t)}{\partial t} = D \nabla^2 \rho_d(\rr,t).
\label{eq:difusion_d}
\end{equation}
If we now assume the Vineyard approximation
\begin{equation}
\rho_d(r,t)= \int\dr'g(\rr')\rho_s(|\rr-\rr'|,t)
\label{eq:vineyard_convol_rho}
\end{equation}
[c.f.\ Eq.\ \eqref{eq:vineyard_convol}], then after Fourier transforming Eq.\
\eqref{eq:difusion_d}, together with Eq.\ \eqref{eq:vineyard_convol_rho}
we obtain
\begin{equation}
\hat{g}(k)\frac{\partial \hat{\rho}_d(k,t)}{\partial t} = -k^2 D\hat{g}(k) \hat{\rho}_d(k,t),
\label{eq:difusion_d_fourier}
\end{equation}
where $\hat{g}(k)$ is the Fourier transform of the radial distribution function $g(r)$
and $\hat{\rho}_d(k,t)$ is the Fourier transform of $\rho_d(r,t)$. Dividing both sides of Eq.\
\eqref{eq:difusion_d_fourier} by $\hat{g}(k)$ and then taking the inverse Fourier
transform we obtain Eq.\ \eqref{eq:diffusion} for $\rho_s(r,t)=G_s(r,t)$.
Thus the Gaussian form in Eq.\ \eqref{eq:Gs_gauss_colloid} together with the Vineyard
approximation \eqref{eq:vineyard_convol_rho} for $\rho_d(r,t)$, together form a self consistent 
solution to the DDFT equations in the dynamical test particle limit, in the case where
we can neglect the contribution from the second term on the right hand side of
Eq.\ \eqref{eq:DDFT_TP}. This term is zero in the ideal-gas limit when the excess
contribution to the free energy $F\subrm{ex}=0$, in Eq.\ \eqref{eq:F}, or when $\rho  \to 0$.
However, we find below for hard spheres
that this approximation is reliable well beyond the ideal gas
regime, which suggests that in the test particle limit $c^{(1)}_s$ and $c^{(1)}_d$
in Eq.\ \eqref{eq:DDFT_TP} must both be slowly varying (almost constant) functions,
so that their gradients are small.

\section{Model for hard spheres}
\label{sec:hard_spheres}
\subsection{Simulation method}

In order to provide benchmark results, we calculate the van Hove function by integrating the equations of motion \eqref{eq:EOM} using standard BD computer simulations \cite{Allen1987}. In order to apply the algorithm we model the hard spheres with a steep continuous pair potential:
\begin{equation} 
\beta v(r)=\left \{ 
\begin{array}{ll}
 (\sigma/r)^{36} -1  \quad &  r<  \sigma, \\
  0 & \mbox{otherwise.} 
\end{array} \right . 
\label{eq:tot}
\end{equation}
We solve Eqs.\ \eqref{eq:EOM} using the Euler forward algorithm using a time step $\delta t= 1 \times 10^{-5} \tau_{B}$; recall that
the Brownian time $\tau_{B}=\sigma^{2}/ D$, where $D=\Gamma k_BT$ is the Stokes-Einstein diffusion coefficient. The random forces  $\zeta_{i}$ in Eq.\ \eqref{eq:EOM} mimic the interaction between particles and solvent, and are sampled from a Gaussian distribution with zero mean and variance $2 D \delta t$.
The simulations are carried out using $N$=1728 particles at densities $\rho  \sigma^3=N(\sigma/L)^{3}=0.2, 0.4, 0.6, 0.8$, and 1 in a cubic box of volume $L^{3}$. 

After an equilibration time of 50 $\tau_{B}$, we sampled the distribution functions  $G_s(r,t)$ and $G_d(r,t)$ at the times $t/\tau_B$=0.01, 0.1, and 1. The distribution functions are averaged over all possible time intervals $t/\tau_B$ of a single simulation run. The total simulated times are $2\tau_B$, $20\tau_B$, and $200\tau_B$ for the short, medium and long time intervals, respectively.
The scaled intermediate scattering function is calculated from the density autocorrelation function in Fourier space, $\phi(k,t)=\langle n_{k}(t) n_{-k}(0) \rangle /\langle n_{k}(0) n_{-k}(0) \rangle $, where $n_k(t)=\sum_{i=1}^N\exp(-i\kk\cdot\rr_i(t))$ are the Fourier components of the local number density.

\subsection{The excess free energy functional}
\label{sec:RY_func}

In order to implement the dynamical test particle
limit we must (as is almost always the case
in density functional theory calculations) select an approximation for the
excess part of the Helmholtz free energy functional, Eq.\ \eqref{eq:F}.
We use the RY functional \cite{ramakrishnan1979fpo},
which is obtained from the two-component generalisation of
Eqs.\ \eqref{eq:Fex_expans} and \eqref{eq:Fex_diff} by neglecting all
terms of order $O(\Delta\rho^3)$ and higher. We obtain:
\begin{align}
F\subrm{ex}&[\rho_s,\rho_d] = Vf\subrm{ex}(\rho ) \notag\\
 &+f'\subrm{ex}(\rho )\left\{\int\dr(\rho_d(\rr)-\rho )+\int\dr\rho_s(\rr)\right\} 
 \notag \\
  & -\frac{1}{2}\int\dr\int\dr'c(|\rr-\rr'|) \bigg\{(\rho_d(\rr)-\rho )(\rho_d(\rr')-\rho ) \notag \\ 
 &\quad +(\rho_d(\rr)-\rho )\rho_s(\rr')+\rho_s(\rr)(\rho_d(\rr')-\rho )\bigg\}
\label{eq:Fex_def}
\end{align}
where $f\subrm{ex}(\rho )$ is the bulk excess free energy per unit volume, $V$ is the volume of the system, $f'\subrm{ex}=\partial f\subrm{ex}/\partial\rho $ and $c(r)$ is the bulk pair direct correlation function of the hard sphere fluid with bulk density $\rho $. We use $f\subrm{ex}$ and $c(r)$ as given by the Percus-Yevick approximation \cite{hansen2006tsl}.

Eq.\ \eqref{eq:Fex_def} is perhaps the simplest functional that one could use to calculate hard sphere fluid density profiles. Our reasons for using this functional are: (i) The structure of the functional is relatively simple (and as a consequence is widely used within liquid state approaches). (ii) Within the RY functional it is straightforward to neglect the species $s$ intra-species interactions which is necessary to ensure that $\rho_s(r,t)$ represents a single particle. (iii) Finally, the RY functional was the first functional to correctly reproduce freezing phenomena in hard spheres.

\subsection{Static structure of the fluid}
\label{sec:static_structure}
\begin{figure}[t]
\centering
\includegraphics[width=8.5cm]{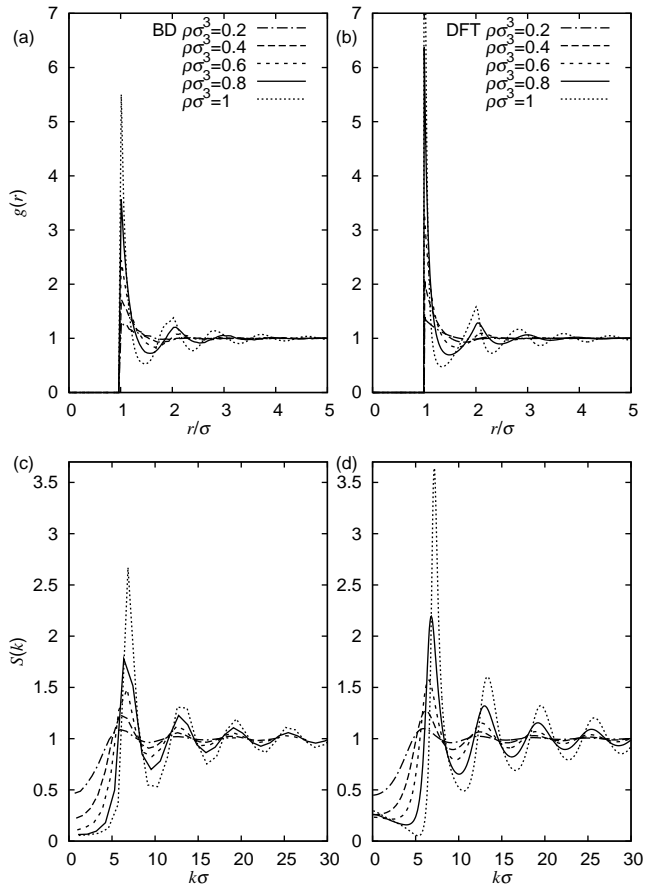}
\caption[Simulation Results.]{\label{fig:equil}
(a) and (b) display radial distribution functions, and (c) and (d) static structure factors for a bulk fluid of hard spheres at densities $\rho \sigma^3=0.2$, 0.4, 0.6, 0.8 and 1. Parts (a) and (c) are obtained from BD simulations and (b) and (d) via Percus' static test particle limit using the RY functional.} 
\end{figure}

In Fig.\ \ref{fig:equil} we display the radial distribution function and static structure factor for a bulk fluid of hard spheres for the densities $\rho \sigma^3=0.2$, 0.4, 0.6, 0.8 and 1. We show results obtained from BD simulations, together with the results from the static test particle limit using the RY approximation for the excess free energy \eqref{eq:Fex_def}. One can observe that for low and intermediate densities $\rho  \sigma^3 \lesssim 0.6$, the test particle results are in good agreement with those from the simulations. However, as the density is increased, the test particle results become less reliable. We see in Fig.\ \ref{fig:equil}(b) that the theory overestimates the contact value, $g(r=\sigma^+)$. This in turn leads to the discrepancies in the static structure factor at high densities in Fig.\ \ref{fig:equil}(d); $S(k)$ is obtained by Fourier transforming the data in (b). The overall conclusion to be drawn from Fig.\ \ref{fig:equil} is that the test particle method combined with the RY functional provides a reliable description of the fluid structure for low and intermediate densities, but at higher densities $\rho  \sigma^3 > 0.6$, the theory is only qualitatively correct.

We return now to the discussion of the somewhat subtle issues concerning the relation between Percus' test particle limit and the zero-dimensionality limit. The results from these two calculations are not the same when one uses an approximate expression for the free energy, such as the RY functional \eqref{eq:Fex_def}. Combining Eqs.\ (\ref{eq:Fex_def}) and (\ref{eq:app_1}), we obtain the following:
\begin{eqnarray}
 \rho_d(r)=\Lambda^{-3} \exp \bigg[ \beta \mu_d-\beta f\subrm{ex}'(\rho )\notag \\
 +\int \dr' c(|\rr-\rr'|)\rho_s(r')\notag \\
 +\int \dr' c(|\rr-\rr'|)(\rho_d(r')-\rho )\bigg].
\label{eq:app_B1}
\end{eqnarray}
Now, recall that a test-particle calculation involves fixing one of the particles at the origin, treating it as an external potential, and then determining the density profile of the fluid (species $d$) under the influence of this external potential. Doing this, using the RY approximation for $F\subrm{ex}[\rho_s,\rho_d]$, Eq.\ \eqref{eq:Fex_def}, we obtain:
\begin{eqnarray}
 \rho_d(r)=\Lambda^{-3} \exp \bigg[ \beta \mu_d-\beta f\subrm{ex}'(\rho )-\beta u(r)\notag \\
 +\int \dr' c(|\rr-\rr'|)(\rho_d(r')-\rho )\bigg].
\label{eq:app_B2}
\end{eqnarray}
Comparing equations (\ref{eq:app_B1}) and (\ref{eq:app_B2}), we see that Eq.\ (\ref{eq:app_B1}) is merely Eq.\ (\ref{eq:app_B2}) with the external potential $\beta u(r)=\beta v(r)$ replaced by the effective potential:
\begin{equation}
\beta u^*(r)=-\int\dr' c(|\rr-\rr'|)\rho_s(r').
\end{equation}
In the $w=0$ limit, when $\rho_s(r)=\delta(r)$, we then obtain:
\begin{equation}
\beta u^*(r)=-c(r).
\label{eq:v_star}
\end{equation}
One consequence of this random phase-like approximation is that the core condition is violated. The degree to which the core condition is violated could be used as an indicator towards the reliability of any approximate free energy functional.

In the remainder of this paper we will display results and distribution functions that are derived from Percus' test particle results used as initial condition, though we will draw attention to results from the zero dimensionality route where appropriate.  

\begin{figure}[t]
\centering
\includegraphics[width=8.5cm]{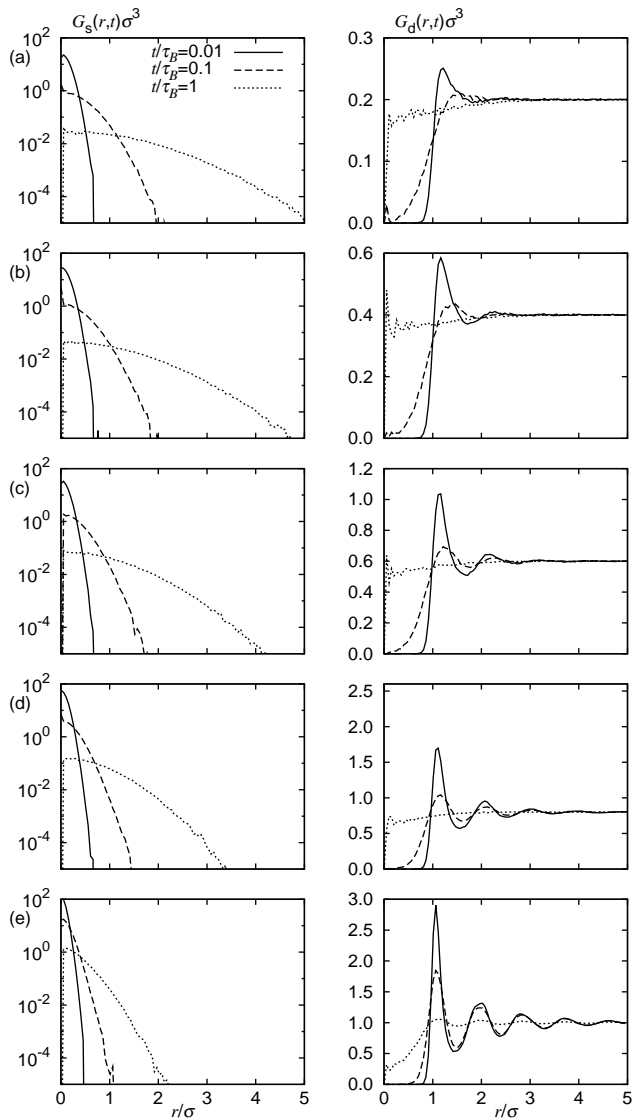}
\caption[Simulation Results.]{\label{fig:sim}
The self and distinct parts of the van Hove distribution function, $G_s(r,t)$ and $G_d(r,t)$, for a hard sphere fluid, measured in BD simulations at densities: (a) $\rho \sigma^3=0.2$ (b) $\rho \sigma^3=0.4$ (c) $\rho \sigma^3=0.6$ (d) $\rho \sigma^3=0.8$ (e) $\rho \sigma^3=1$. The results are plotted for times $t/\tau_B$=0.01 (solid line), 0.1 (dashed line) and 1 (dotted line).  In the semi-logarithmic scale of $G_s(r,t)$ versus $r$ a Gaussian appears as a parabola. The $G_d(r,t)$ results are shown on a linear scale.} 
\end{figure}

\begin{figure}[t]
\centering
\includegraphics[width=8.5cm]{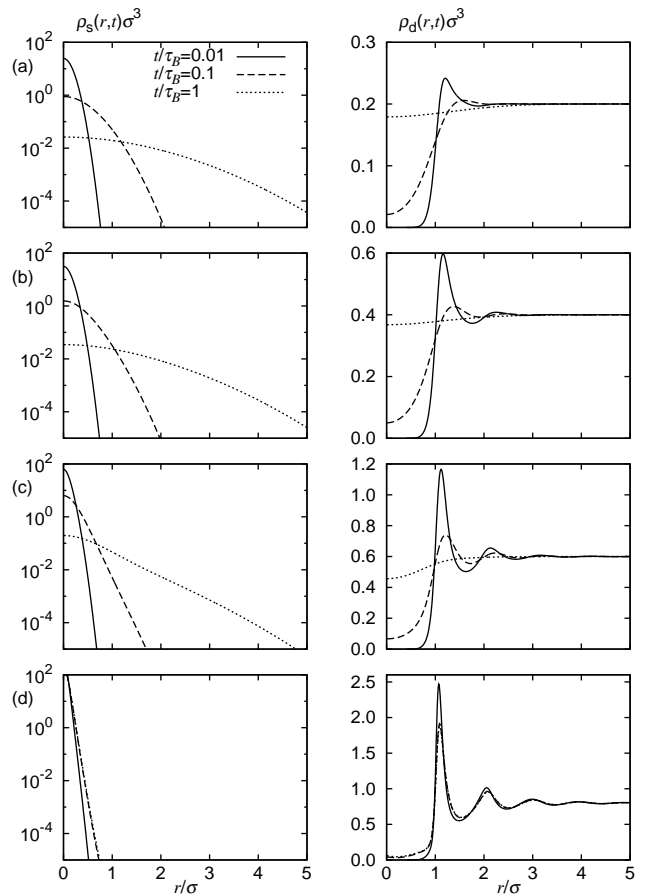}
\caption[DDFT Results.]{\label{fig:ddft}
The `$s$' and `$d$' density profiles, $\rho_s(r,t)$ and $\rho_d(r,t)$, obtained from the dynamical test particle theory, for densities: (a) $\rho \sigma^3=0.2$ (b) $\rho \sigma^3=0.4$ (c) $\rho \sigma^3=0.6$ (d) $\rho \sigma^3=0.8$. The results are plotted for times $t/\tau_B$=0.01 (solid line), 0.1 (dashed line) and 1 (dotted line). In (d), after a short time, the system reaches an `arrested state', where the density profiles no longer evolve in time and the width of $\rho_s(r,t \to \infty)$ remains finite.} 
\end{figure}

\begin{figure}[t]
\centering
\includegraphics[width=8.5cm]{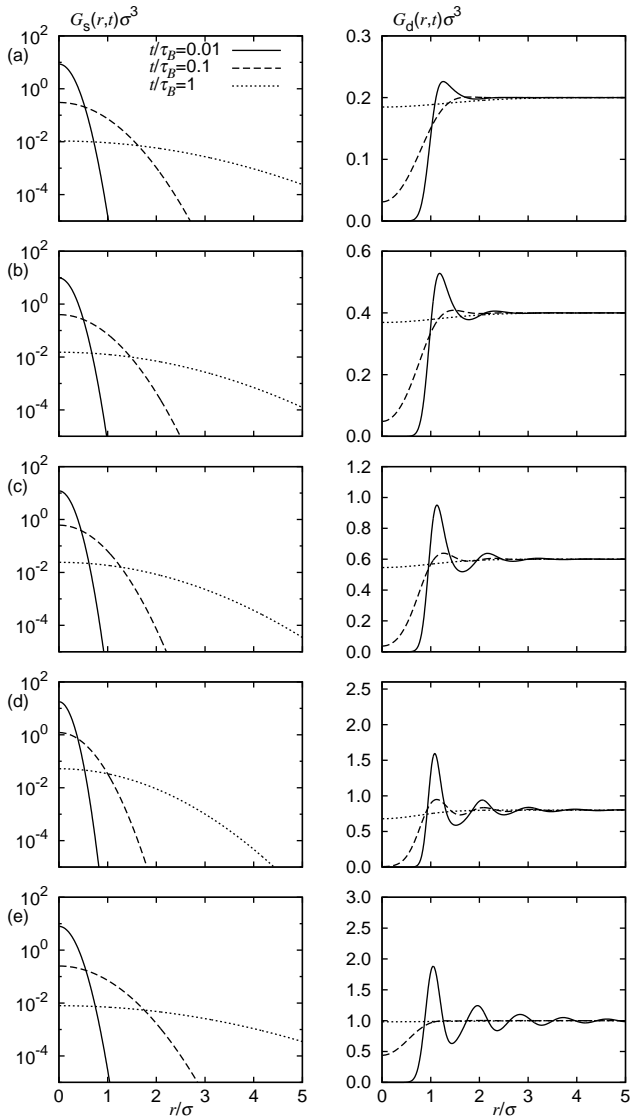}
\caption[Simulation Results.]{\label{fig:vineyard}
The self and distinct parts of the van Hove distribution function, $G_s(r,t)$ and $G_d(r,t)$, calculated using the Vineyard approximation for densities: (a) $\rho \sigma^3=0.2$ (b) $\rho \sigma^3=0.4$ (c) $\rho \sigma^3=0.6$ (d) $\rho \sigma^3=0.8$ (e) $\rho \sigma^3=1$. The results are plotted for times $t/\tau_B$=0.01 (solid line), 0.1 (dashed line) and 1 (dotted line).} 
\end{figure}

\begin{figure}[t]
\centering
\includegraphics[width=8.5cm]{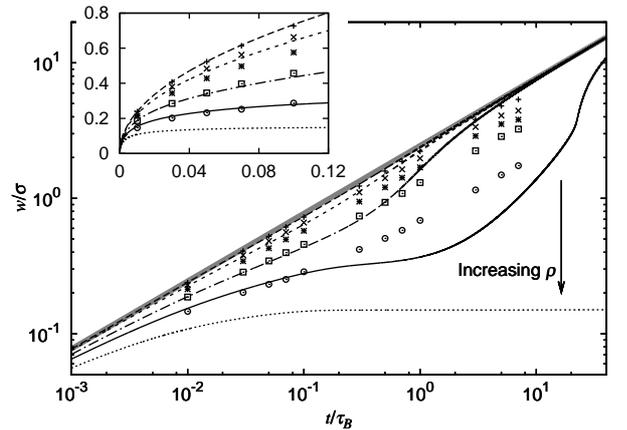}
\caption[DDFT Results.]{\label{fig:width_time}
The width $w$ of the self part of the van Hove function, defined in Eq.\ \eqref{eq:width}, measured in the BD simulations for $\rho \sigma^3=0.2$ ($+$), 0.4 ($\times$), 0.6 ($\ast$), 0.8 ($\Box$), and 1 ($\medcirc$), and from the dynamical test particle theory, for $\rho \sigma^3=0.2$ (long dashed line), 0.4 (short dashed line), 0.6 (dashed-dotted line), 0.7 (solid line), and 0.8 (dotted line). Also shown is the result from the Vineyard approximation (thick solid grey line). In the main panel both axes are logarithmic; the inset displays the same results on linear scales.} 
\end{figure}

\begin{figure}[t]
\centering
\includegraphics[width=8.5cm]{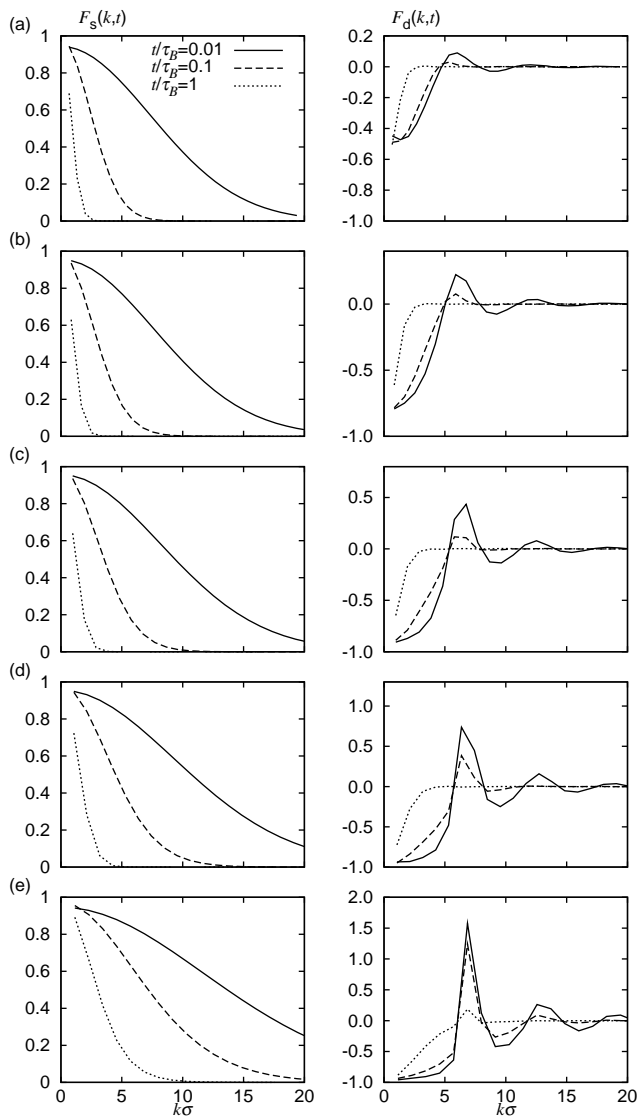}
\caption[Simulation Results.]{\label{fig:sim_ft}
Intermediate scattering function $F(k,t)$ as a function of the scaled wave vector $k\sigma$, obtained by a spatial Fourier transform, Eq.\ \eqref{eq:FT}, of the BD simulation results for the van Hove functions displayed in Fig.~\ref{fig:sim}.} 
\end{figure}

\begin{figure}[t]
\centering
\includegraphics[width=8.5cm]{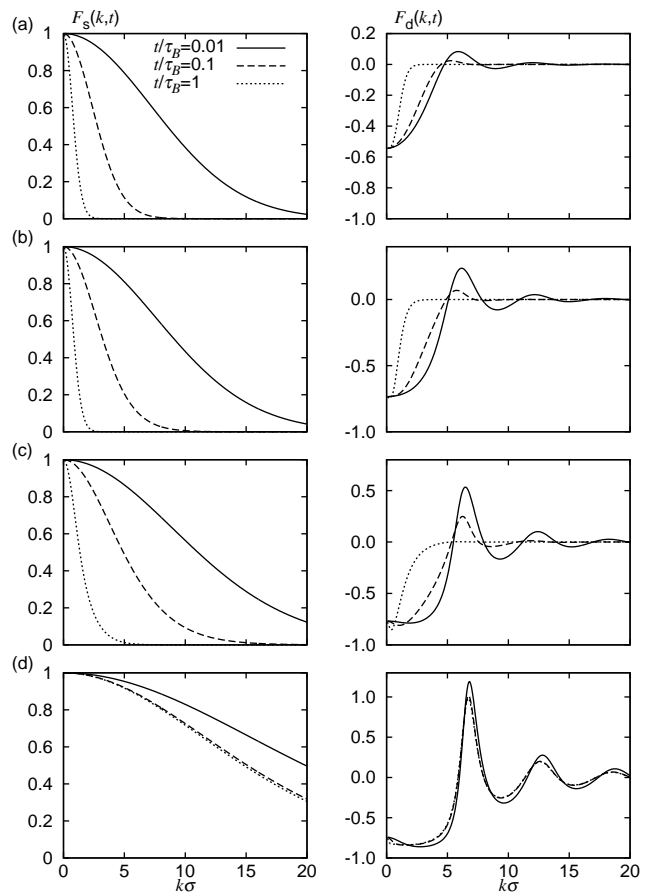}
\caption[DDFT Results.]{\label{fig:ddft_ft}
Intermediate scattering functions $F(k,t)$, obtained by a spatial Fourier transform, Eq.\ \eqref{eq:FT}, of the dynamic test particle density profiles, $\rho_s(r,t)$ and $\rho_d(r,t)$, displayed in Fig.~\ref{fig:ddft}.} 
\end{figure}

\begin{figure}[t]
\centering
\includegraphics[width=8.5cm]{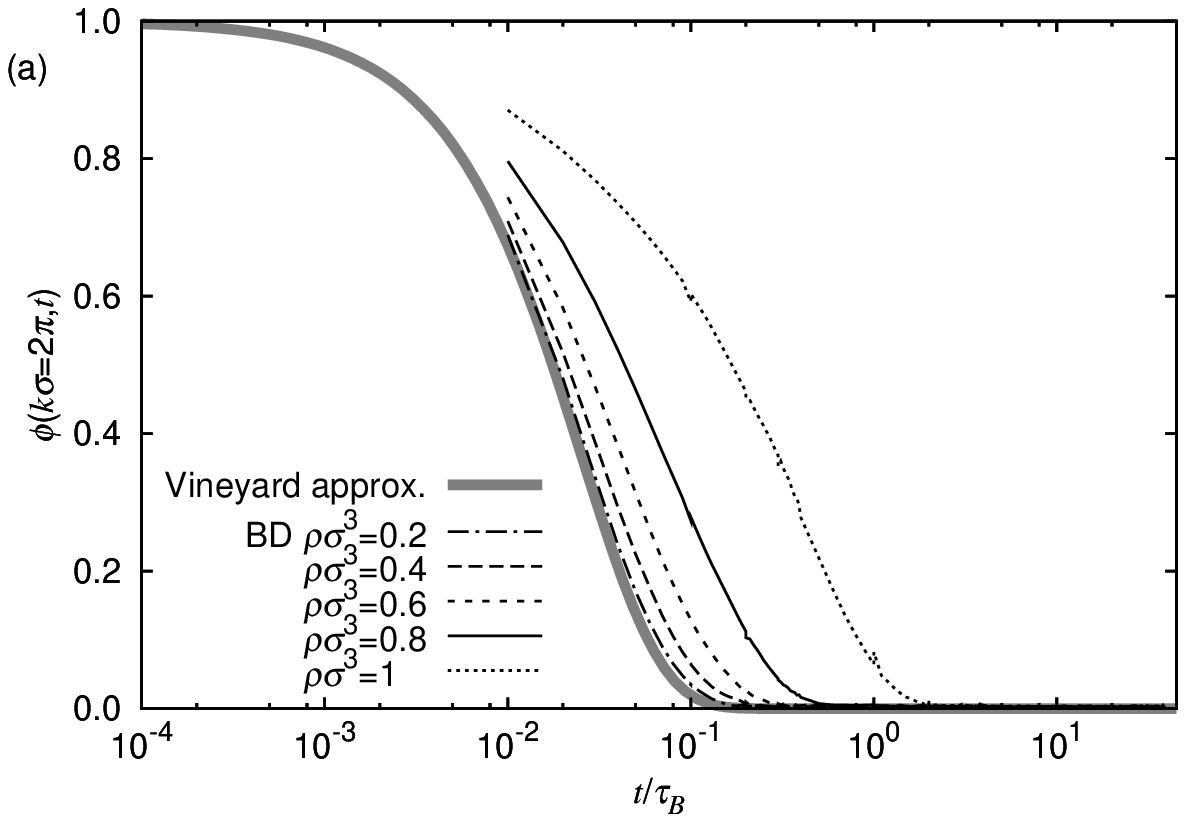}
\includegraphics[width=8.5cm]{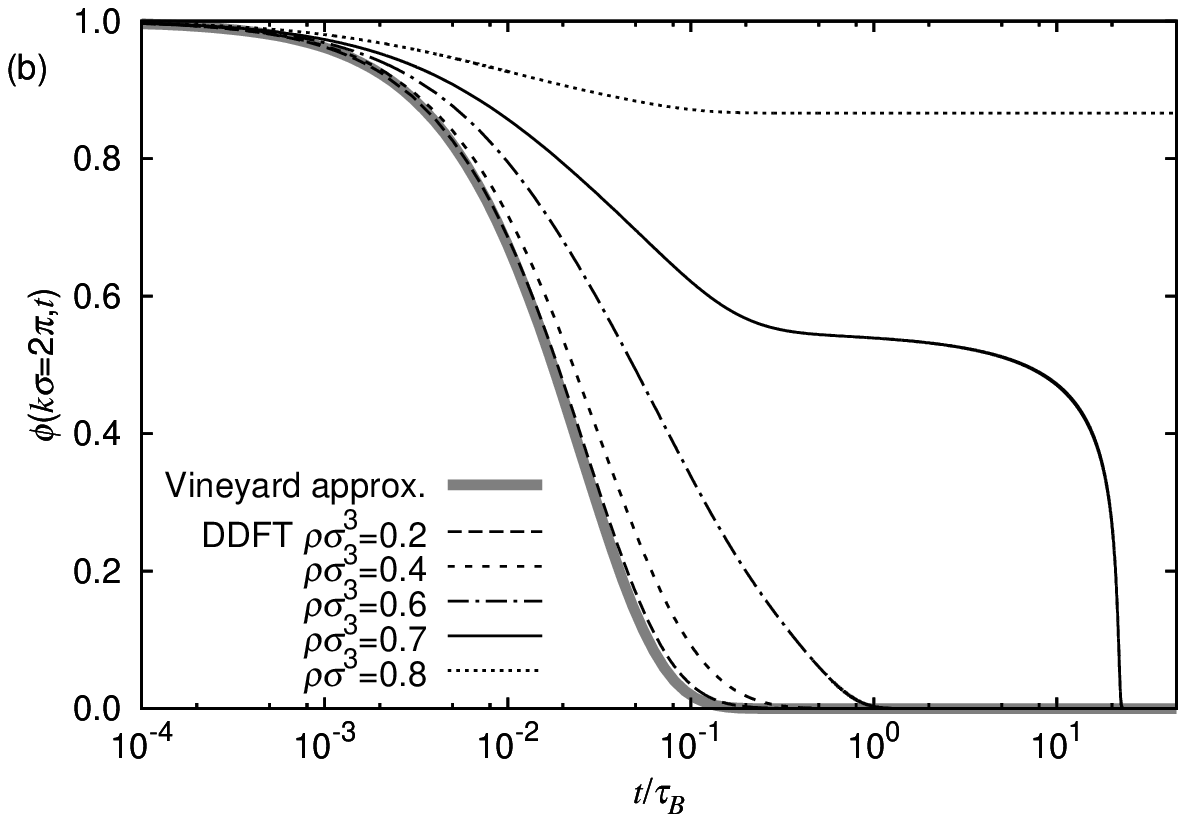}
\caption[DDFT Results.]{\label{fig:intermediate}
Scaled intermediate scattering function $\phi(k\sigma=2\pi,t)$ as a function of time $t/\tau_B$, calculated from the Vineyard approximation (thick gray line), compared to (a) BD simulation results, and (b) the dynamical test particle method. The dynamical test particle results exhibit slowing and arrested dynamics for $\rho \sigma^3=0.7$ and 0.8.}
\end{figure}

\section{Results}
\label{sec:results}
\subsection{Dynamic approaches}

In Fig.\ \ref{fig:sim} we display the two parts of the van Hove functions, $G_s(r,t)$ and $G_d(r,t)$, measured in the BD simulations for fluid densities $\rho \sigma^3=0.2$, 0.4, 0.6, 0.8 and 1. The different curves correspond to the times $t/\tau_B=0.01$, 0.1, and 1. $G_s(r,t)$ appears to have a near-Gaussian form for all times and densities. For short times $G_d(r,t)$ exhibits a correlation hole for $r<\sigma$ and it is highly structured for (larger) $r>\sigma$. At later times the structure in $G_d(r,t)$ diminishes and the correlation hole becomes `filled in'. Recall that Eq.\ \eqref{eq:van_hove_7} defines the long time limit. Increasing the density beyond $\rho \sigma^3=1$, we find that the simulated system crystallises onto a regular lattice and there is no evidence of glass forming behaviour.

In Fig.\ \ref{fig:ddft} we display the one-body density profiles, $\rho_s(r,t)$ and $\rho_d(r,t)$, from the DDFT dynamical test particle method for bulk fluid densities $\rho \sigma^3=0.2$, 0.4, 0.6 and 0.8. As initial condition, $\rho_d(r,t=0)=g(r)$, we have used Percus' test particle method for calculating $g(r)$, as shown in Fig.~\ref{fig:equil}. The results in Fig.\ \ref{fig:ddft} correspond to the same times as the BD curves displayed in Fig.~\ref{fig:sim}, namely $t/\tau_B=0.01$, 0.1, and 1. Comparing the BD results in Fig.\ \ref{fig:sim} with the DDFT results in Fig.~\ref{fig:ddft} we observe that for densities $\rho \sigma^3=0.2$, 0.4, and 0.6, there is good qualitative agreement between the simulation and DDFT results. The $\rho_d(r,t)$ results show a similar amount of structure as the $G_d(r,t)$ results, and $\rho_s(r,t)$ has a very similar magnitude and range as $G_s(r,t)$, although for $\rho \sigma^3=0.6$, $\rho_s(r,t)$ shows some departures from the almost Gaussian shape observed in the simulation results, particularly at $t/\tau_B=1$. For $\rho \sigma^3=0.8$ we find that the dynamic test particle method predicts that after a short time $t/\tau_B\sim 0.1$ the density profiles $\rho_s(r,t)$ and $\rho_d(r,t)$ cease to change with time and that the system becomes `arrested'. One could interpret this state as the tagged particle remaining localised within the cage formed by the neighbouring fluid particles. We discuss the significance of this phenomenon in Sec.~\ref{sec:conclusion}.

In Fig.\ \ref{fig:vineyard} we show the van Hove functions calculated using the Vineyard approximation, Eqs.~\eqref{eq:Gs_gauss_colloid} and \eqref{eq:vineyard_convol} with $D_l=D=k_BT\Gamma$, for fluid densities $\rho \sigma^3=0.2$, 0.4, 0.6, 0.8 and 1 and times $t/\tau_B=0.01$, 0.1, and 1. As in the DDFT we have used $g(r)$ calculated using the RY functional and Percus' test particle method, although one could use $g(r)$ obtained from any reasonable method, including $g(r)$ measured in the BD simulations. Comparing the Vineyard results to the BD simulation results in Fig.\ \ref{fig:sim}, we find that there is reasonably good agreement between the two. The form of $G_s(r,t)$ is fixed to be Gaussian, so there is good agreement with $G_s(r,t)$ from the BD simulations, though it is clear that the width of $G_s(r,t)$ increases more rapidly in the Vineyard approximation. For densities $\rho \sigma^3 \leq 0.8$ there is a similar amount of structure present in $G_d(r,t)$ for $r>\sigma$ in the Vineyard approximation as in the simulation results. However, for $\rho \sigma^3=1$ the Vineyard approximation does not exhibit the same degree of structure that is present in the simulation results. 

In Fig.~\ref{fig:width_time} we compare the width, $w(t)$, of the self part of the van Hove function, $G_s(r,t)$, obtained from (i) BD simulation results, (ii) dynamical test particle limit, and (iii) the Vineyard approximation. In the Vineyard approximation the time dependence of $w(t)$ is defined by Eq.\ \eqref{eq:diff_atom} and does not depend on density, so there is only one master curve. This is because in the same way as in the dynamical test particle limit, we set $D_l=D$, where $D=k_BT \Gamma$ is the short time diffusion coefficient, which is strictly only equal to the long time self diffusion coefficient $D_l$ in the limit $\rho  \to 0$. Since $w(t)\propto\sqrt{t}$, on the double logarithmic scale in Fig.\ \ref{fig:width_time} this is represented by a straight line with gradient $1/2$. We find that the simulation results are also approximately linear in this representation for all densities considered, but that there is a slowing down effect as density is increased, due to the fact that $D_l$ decreases as the fluid density is increased and is no longer equal to $D$. For $\rho\sigma^3=0.2$, the simulation results are close to the Vineyard result. As the bulk density is increased, the BD results move away from this line.

The dynamical test particle results for $w(t)$ in Fig.~\ref{fig:width_time} exhibit a much stronger dependence on density. At low densities the curves are similar to the Vineyard and the BD results, but as the density is increased the $w(t)$ curves show a slowing down, and then (unphysical) speeding up of the dynamics, unlike that seen in the BD results. This slowing down is greatly exaggerated so that the DDFT curve for $\rho \sigma^3=0.6$ is similar to the BD result at $\rho \sigma^3=0.8$. Furthermore, the DDFT curves for $\rho \sigma^3=0.7$, and 0.8 have no counterpart in the simulation results. For $\rho \sigma^3=0.7$ the $w(t)$ curve shows extremely exaggerated slowing down and speeding up. We believe that the unphysical speeding up for $t/\tau_B \gtrsim 10^1$ is due to the fact that the DDFT incorrectly sets the long-time diffusion coefficient $D_l$ equal to the short time diffusion coefficient $D$, so that as the particle escapes the cage of neighboring particles, it is forced to ``catch-up'' to give the incorrect long time behavior. Note that from the Smoluchowski
equation (22) it can be shown\cite{Nageleetal} that $w(t)$ must be sub-diffusive for intermediate times, and that
the long time diffusion coefficient must be smaller than the short time
one, a feature which is well
established in Brownian dynamics simulations and experiments\cite{Brady1, Brady2, FossBrady,Weeksetal}. The curve for $\rho \sigma^3=0.8$ shows that the system slows down so much that the dynamics are arrested, so that $w(t \to \infty)$ is finite, as one would infer from the density profiles shown in Fig.~\ref{fig:ddft}. We postpone a discussion of the possible physical implications to Sec.~\ref{sec:conclusion}.

For completeness, we also plot the intermediate scattering function $F(k,t)$. In Fig.~\ref{fig:sim_ft} we display results from BD computer simulations, and in Fig.~\ref{fig:ddft_ft} the results from the DDFT. We find that for $\rho \sigma^3=0.2$, 0.4 and 0.6, the results from both approaches exhibit very similar structure. However, since in the DDFT the dynamics become arrested at $\rho \sigma^3=0.8$, so $F(k,t)$ becomes arrested after a very short time, unlike the BD simulations result. In Fig.~\ref{fig:intermediate} we plot the scaled intermediate scattering function,
\begin{equation}
\phi(k,t)=\frac{F_s(k,t)}{F_s(k,t=0)},
\label{eq:isf}
\end{equation}
for fixed $k\sigma=2\pi$, obtained from the Vineyard approximation and compare to the BD simulation results (Fig.\ \ref{fig:intermediate}(a)) and the DDFT results (Fig.\ \ref{fig:intermediate}(b)). At the lower densities the BD simulation results and the DDFT results are close to the Vineyard approximation and both show some slowing down with density. At the higher densities the BD results continue to show a steady decay. However, in the DDFT results for $\rho \sigma^3=0.7$ we see $\phi(k\sigma=2\pi,t)$ decays in two stages over a much longer time. For $\rho \sigma^3=0.8$ the arrested dynamics cause $\phi(k\sigma=2\pi,t)$ to remain finite in the limit $t\to\infty$.

\subsection{Relating dynamic to static density profiles}

\begin{figure}[t]
\centering
\includegraphics[width=8.5cm]{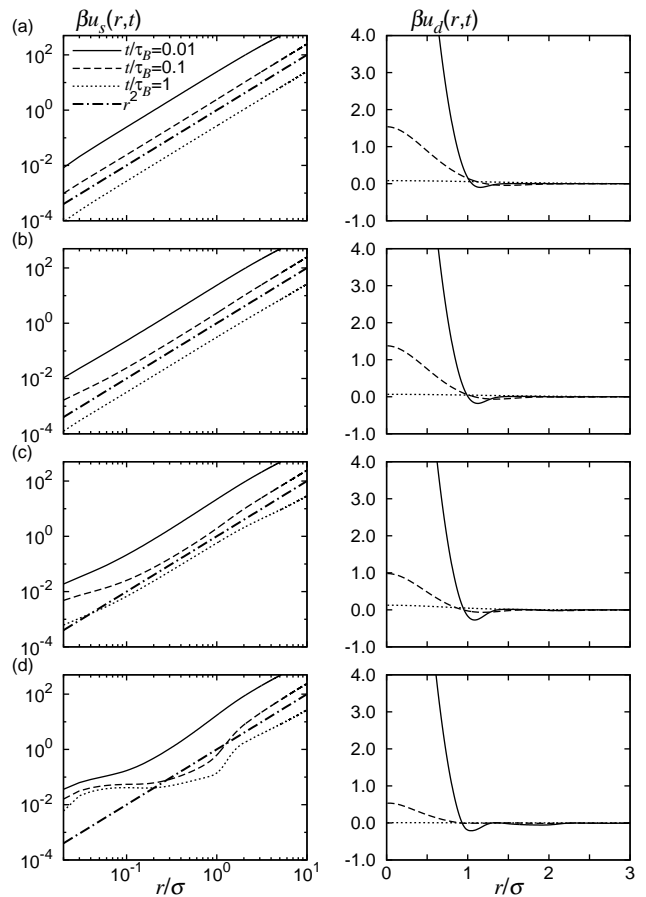}
\caption[DDFT Results.]{\label{fig:ddft_v}
External potentials, $u_s(r,t)$ and $u_d(r,t)$, required to generate equilibrium density profiles, $\rho_s(r,t)$ and $\rho_d(r,t)$, identical to those obtained from the dynamical test particle approach -- see Fig.~\ref{fig:ddft}. The external potentials also include an overall time-dependent additive constant which is not indicated. Note the logarithmic $x$-axis for $u_s(r,t)$. At low densities $u_s(r,t)$ is approximately parabolic, as indicated by the straight line (dashed-dotted). At high densities $u_s(r,t)$ is distorted at small $r$, but still parabolic at large $r$.} 
\end{figure}

One may connect DDFT and equilibrium DFT by finding the unique set of effective external potentials $\{u_i(\rr)\}$ that in equilibrium generate the same set of density profiles as obtained in the dynamic approach at a particular time $t$. These potentials represent the net effect of finite time and limited diffusion preventing the fluid from finding the structure that minimises the system free energy. For the two-component system considered here, the two external potentials $\beta u_i(r,t)$ may be recovered, up to an overall time-dependent additive constant $\beta\mu_i(r,t)$, by rearranging Eq.~\eqref{eq:app_1}:
\begin{equation}
\beta u_i(r,t)-\beta\mu_i=c_i^{(1)}(r;[{\rho_s,\rho_d}])-\ln[\Lambda^3\rho_i(r,t)],
\label{eq:vext_el}
\end{equation}
where $\rho_s(r,t)$ and $\rho_d(r,t)$ are the solution of the DDFT at time $t$. In Fig.~\ref{fig:ddft_v} we plot these external potentials corresponding to the density profiles from DDFT displayed in Fig.~\ref{fig:ddft}. We find that at the lowest densities $\rho \sigma^3=0.2$ and $0.4$, the shape of $u_s(r,t)$ is approximately parabolic for all times and distances $r$. As the fluid density is increased, $u_s(r,t)$ departs from the parabolic form. For $\rho \sigma^3=0.8$ the curves are still parabolic at large $r$, but at small $r$ they become distorted. We find that $u_d(r,t)$ does not vary significantly with density. At short times it is dominated by strong repulsion within the hard-core diameter, $r<\sigma$. Recall that in order to calculate $g(r)$, which corresponds to $t=0$, we chose to use Percus' test particle method, and hence have introduced an external potential equal to the hard sphere potential. The strong repulsion found for short times is a remnant of this external potential. As $t$ increases, the strength of this repulsion decreases and becomes almost zero for $t/\tau_B=1$~\cite{footnote2}.

\subsection{Corresponding equilibrium approach}

Having established the form of the external potentials necessary to create equilibrium fluid density profiles equal to the profiles calculated using the dynamical test particle method, we now seek a simple approximation for this set of external potentials, to allow us to easily calculate equilibrium density profiles that mimic the dynamic profiles. In other words, we seek to determine the full van Hove function when $G_s(r,t)$ has a given width $w$, without calculating the entire preceding time series of profiles. In doing this we lose time $t$ as a function argument and instead we must `label' the density profiles with $w$. In what follows, we will disregard the associated problem of relating $w$ to time $t$.

We parametrise the external potentials using a simple functional form. Firstly, we assume that $u_s(r,w)$ is parabolic for all widths:
\begin{equation}
\beta u_s(r,\lambda)=\lambda r^2,
\label{eq:us_para}
\end{equation}
where $\lambda$ is the strength of the confining potential, and $w$ is now an unknown function of $\lambda$. For the external potential that acts on species $d$ we consider two options. The first is to assume that $u_d(r,w)=0$. In this case, it is possible to solve exactly for the equilibrium distribution functions and the free energy, as outlined in Appendix~\ref{app:A}. We find that the species $s$ density profile is a Gaussian,
\begin{equation}
\rho_s(r,\lambda)=\frac{\exp(-\beta \lambda r^2)}{(\pi/\beta\lambda)^{3/2}},
\label{eq:exact_s}
\end{equation}
where the dependence of the width $w$ on $\lambda$ is,
\begin{equation}
w(\lambda) = (2\lambda/3)^{-1/2},
\label{eq:w_lambda}
\end{equation}
and the $d$ profile is given by a convolution of the radial distribution function $g(r)$, together with the Gaussian profile $\rho_s(r)$, and multiplied by $\rho $:
\begin{equation}
\rho_d(r) =\rho \int\drr' \rho_s(\rr')g(|\rr-\rr'|).
\label{eq:gd_conv}
\end{equation}
These distribution functions are identical to those from the (dynamic) Vineyard approach.

The second approach that we consider is to calculate the density profiles without defining the external potential $u_d(r,w)$ at the outset of the calculation. Instead, we determine this potential self consistently `on-the-fly' as part of our iterative numerical solution routine, based on the following considerations: Firstly, recall the normalisation constraints on the van Hove function in Eqs.\ \eqref{eq:van_hove_4.5} and \eqref{eq:van_hove_5}. In order to satisfy the normalisation constraint  \eqref{eq:van_hove_4.5} on the density profile for the single tagged $s$ particle, we introduce a Lagrange multiplier $\mu_s$. One may also consider $\lambda$ to be a Lagrange multiplier that enforces the width constraint \eqref{eq:width} on the profile $\rho_s(r)$. In our calculations the value of $\mu_s$ is determined on-the-fly by enforcing Eq.\ \eqref{eq:van_hove_4.5} at each step of our iterative routine. However, we are not able to do the same for the density profile of the remaining $d$ particles, because we also must have
\begin{equation}
\rho_d(r,w)\to\rho , \ {\rm as} \ r \to\infty.
\label{eq:rho_d_lim}
\end{equation}
Multiplying $\rho_d(r)$ by a single factor breaks this condition, so we are not able to simply enforce \eqref{eq:van_hove_5} at each step of our iterative routine in the same way as we do for $\rho_s(r)$. The condition in Eq.\ \eqref{eq:rho_d_lim} implies that we require an {\it a priori} unknown inhomogeneous external potential, $u_d(r,w)$, with the property that $u_d(r,w)\to0$ as $r\to\infty$. This may be achieved by scaling the quantity $\rho_d(r)-\rho $ (instead of scaling $\rho_d(r)$ itself) at each step, so that $\rho_d(r)$ satisfies both \eqref{eq:van_hove_5} and \eqref{eq:rho_d_lim}. Once convergence of the numerical procedure is achieved, one may then inspect the effective external potential $u_d(r,w)$ by substituting the resulting density profiles into Eq. \eqref{eq:vext_el}.

\begin{figure}[t]
\centering
\includegraphics[width=8.5cm]{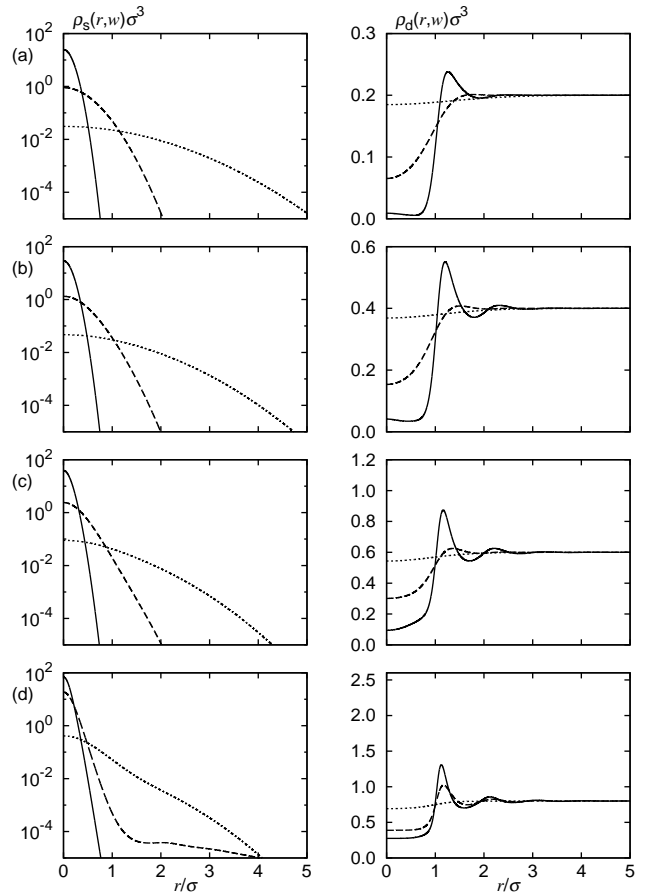}
\caption[DFT Results.]{\label{fig:dft}
The density profiles $\rho_s(r,w)$ and $\rho_d(r,w)$, calculated using the equilibrium DFT. The curves are calculated at the densities: (a) $\rho \sigma^3=0.2$, (b) $\rho \sigma^3=0.4$, (c) $\rho \sigma^3=0.6$ and (d) $\rho \sigma^3=0.8$. The curves are chosen so that the widths $w$ of $\rho_s(r,w)$ correspond to the same widths of $G_s(r,t)$ at times $t/\tau_B$=0.01 (solid line), 0.1 (dashed line) and 1 (dotted line) displayed in Fig.~\ref{fig:sim}.} 
\end{figure}

The density profiles calculated using this equilibrium method are shown in Fig.~\ref{fig:dft} where we plot $\rho_d(r,w)$ and $\rho_s(r,w)$ having widths identical to those of the van Hove functions from simulation, displayed in Fig.~\ref{fig:sim}. Note that we consider only the densities, $\rho \sigma^3=0.2$, 0.4, 0.6 and 0.8. These equilibrium profiles have been calculated using a normalisation constant taken from the approximation for $g(r)$ calculated using Percus' test particle method. We find that there is reasonable qualitative agreement between the equilibrium DFT density profiles displayed in Fig.\ \ref{fig:dft} and the BD simulation results displayed in Fig.~\ref{fig:sim}. However, the profiles predict too much infilling in the region close to the origin $r<\sigma$, particularly at higher densities, which in turn results in an underestimate in the structure of the profiles at larger $r>\sigma$. This error occurs for the reasons discussed in the previous subsection \ref{sec:static_structure}; i.e.\ that the RY functional does not exert a strong enough interaction from the test particle onto the rest of the fluid. 

For $\rho \sigma^3=0.8$, shown in Fig.\ \ref{fig:dft}(d), we are able to calculate density profiles for all values of $w$, even though in the DDFT the profiles became `trapped' at small $w$. The most striking aspect of these density profiles is that for intermediate values of $w$ we see that $\rho_s(r)$ exhibits a plateau and a long tail. These features were not observed in the BD simulation results. However, similar features are present in $G_s(r,t)$ at intermediate times for colloidal spheres at densities close to the glass transition, where they are interpreted as the signature of dynamical heterogeneity in the system \cite{kegel2000dod}.

\subsection{Free energy landscape}

\begin{figure}[t]
\centering
\includegraphics[width=8.5cm]{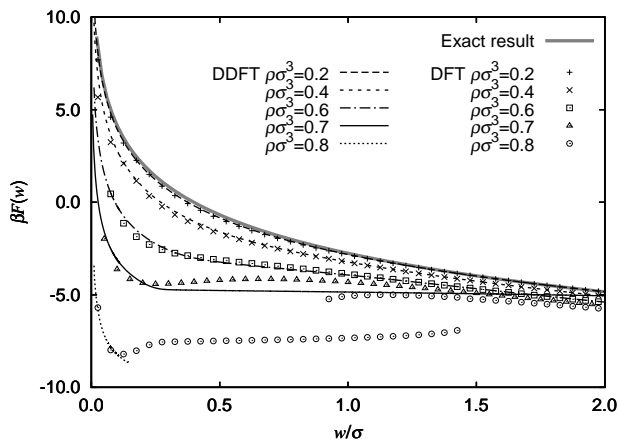}
\caption[Energy landscape.]{\label{fig:energy_landscape}
The free energy landscape $F(w)$ plotted as a function of $w$, the width of $\rho_s(r,w)$, calculated using both the DDFT and equilibrium DFT approaches, compared to the exact result, Eq.\ \eqref{eq:Fw_exact}. For $\rho \sigma^3=0.8$ we find two disconnected branches of $F(w)$.} 
\end{figure}

\begin{figure}[t]
\centering
\includegraphics[width=8.5cm]{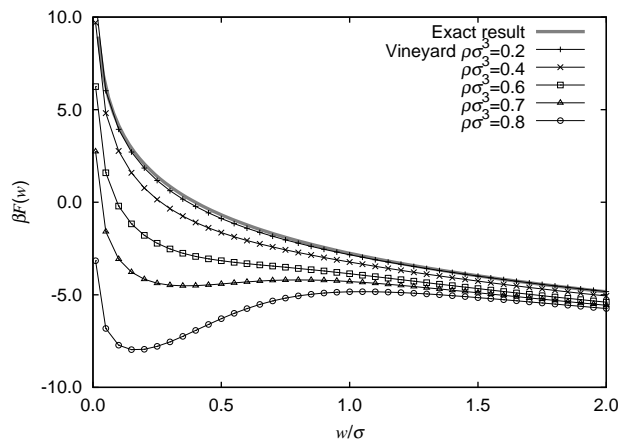}
\caption[Vineyard energy landscape.]{\label{fig:vineyard_energy_landscape}
The free energy landscape $F(w)$ plotted as a function of the width $w/\sigma$, calculated by substituting the density profiles from the exact equilibrium solution  [Eqs.\ \eqref{eq:exact_s} and \eqref{eq:gd_conv}] into the RY functional. Here the deviation from the exact free energy and the emergence of the minimum are entirely due to the RY functional.} 
\end{figure}

\begin{figure}[t]
\centering
\includegraphics[width=8.5cm]{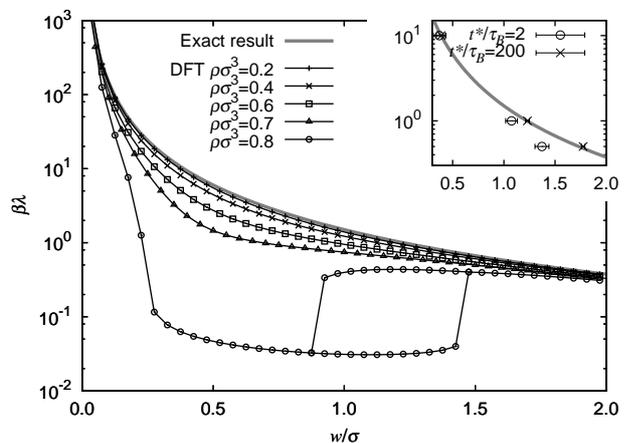}
\caption[DFT Results.]{\label{fig:width_lambda}
The (scaled) strength of the parabolic potential, $\beta\lambda$, versus the resulting width, $w/\sigma$, of $\rho_s(r)$. Results are calculated using the equilibrium DFT and compared to the exact result. For the bulk fluid densities $\rho \sigma^3=0.2$, 0.4, 0.6 and 0.7, $\lambda$ decreases monotonically with $w$. However, for $\rho \sigma^3=0.8$ the curve is no longer monotonic, and exhibits two disconnected branches. The inset shows the effect of insufficient simulation run times on the equivalent situation studied using BD computer simulations, where $t^*$ is the simulation run time.} 
\end{figure}

Since we are able to convert the dynamic density profiles into their equilibrium equivalents via a set of external potentials, c.f. Eq.~\eqref{eq:vext_el}, we are also able to calculate the equilibrium Helmholtz free energy for this corresponding equilibrium situation. Although this free energy is strictly an equilibrium quantity, since it underlies the time evolution of our dynamic approach we believe that it plays a relevant role. Therefore, by substituting the density profiles calculated using the DDFT into the free energy functional, Eqs.\ \eqref{eq:F} and~\eqref{eq:Fex_def}, we are able to map out a `free energy landscape' as a function of $t$ or $w$. Fig.\ \ref{fig:energy_landscape} plots this free energy landscape, $F(w)$, for densities $\rho \sigma^3=0.2$, 0.4, 0.6, 0.7 and 0.8. For $\rho\sigma^3=0.2$, 0.4, and $0.6$ we find that $F(w)$ decreases monotonically with $w$. This decrease is initially steep and then the gradient begins to reduce as $w$ increases. For $\rho\sigma^3=0.7$, we find that after the initial steep descent the landscape develops an almost constant plateau, but there is still a very small negative gradient. For $\rho\sigma^3=0.8$ the decrease is rapid and then the landscape terminates abruptly as the density profiles reach an arrested state.

In the equilibrium case where $u_d(r,w)=0$, one has an exact solution (see Appendix \ref{app:A}) for the free energy landscape as a function of the width,
\begin{equation}
F(w) = F_{\rm id} - \ln(Z'_N)-\frac{3}{2}\ln\left(\frac{2\pi w^2}{3\beta}\right),
\label{eq:Fw_exact}
\end{equation}
where $F_{\rm id}$ is the ideal Helmholtz free energy, and $Z'_N$ is an irrelevant constant representing the partition function of the fluid when the test particle is located at the origin. We plot Eq.\ \eqref{eq:Fw_exact} alongside the landscapes from the DDFT approach in Fig.~\ref{fig:energy_landscape}. We find that this curve is located close to the DDFT landscape for $\rho \sigma^3=0.2$ but that the deviation grows with increasing density.

We also calculate the free energy landscape via the equilibrium approach described in the previous subsection and compare this to the results from both the dynamic approach and the exact equilibrium result in Fig.~\ref{fig:energy_landscape}. For the lower densities, $\rho \sigma^3=0.2$, 0.4 and 0.6, we find that there is good agreement between the DDFT and equilibrium DFT approaches. For $\rho \sigma^3=0.7$ we find that there is good agreement at low $w$, but around $w/\sigma=0.7$ there is a local maximum in $F(w)$ in the equilibrium DFT results that is not present in the DDFT results. For $\rho \sigma^3=0.8$ we find that the DDFT free energy landscape terminates abruptly at a fairly low value of $w$. However, since we can calculate the density profiles using our equilibrium approach for all widths we can therefore calculate $F(w)$ for all $w$. We find a free energy landscape with two disconnected branches. Therefore, for a range of values of $\lambda$ we find two solutions to the Euler-Lagrange equations with different widths. Whether this is an indication of dynamic heterogeneity or an artefact of the functional is an interesting question. 

If we take the density profiles calculated using the exact route [Eqs.\ \eqref{eq:exact_s} and \eqref{eq:gd_conv}] and substitute these into the RY functional, we find that the free energy curves, shown in Fig.~\ref{fig:vineyard_energy_landscape}, do not follow the exact result \eqref {eq:Fw_exact}, but are very similar to those from the DDFT and DFT approaches and even exhibit minima for $\rho \sigma^3=0.7$ and 0.8. Therefore, we must conclude that it is largely a property of the RY approximation for the free energy functional that generates these minima.

In Fig.~\ref{fig:width_lambda} we plot the exact relationship, Eq.~\eqref{eq:w_lambda}, between the strength of the confining potential, $\lambda$, against the width, $w$, and compare it to the results from the equilibrium DFT approach. At the lowest densities the equilibrium DFT approach closely follows the exact result, but as the density is increased the width decreases for a given $\lambda$. For $\rho \sigma^3=0.8$ we find that this curve is no longer monotonic and has two disconnected branches.

Systems which are in a glassy or jammed state are by definition non-ergodic. A way of modelling non-ergodicity in Brownian dynamics is to measure the state of the system over too short a time frame. We demonstrate this effect by simulating a fluid where a single tagged particle is trapped in a parabolic potential well; c.f. Eq.~\eqref{eq:us_para}. If the radius of the well is sufficiently large and the simulation time is too small, then the particle is not able to fully explore the outer regions of the potential well. In the inset of Fig.~\ref{fig:width_lambda} we plot two results pertaining to this scenario where the fluid density is $\rho \sigma^3=0.8$, and $t^*$ is the simulation run time. The results for the longer simulation run time agree well with the exact result, but results over the shorter time underestimate the width, particularly at the smaller  values of $\lambda$. A similar effect may exist in the DFT approach, where the non-ergodicity arises from not including the states where the particle is far from the origin. Recall that formally the density profile from DFT is an average over all possible states of the system. Using an approximate functional some of these states may be neglected \cite{reguera:2558}.

\section{Concluding Remarks}
\label{sec:conclusion}

On the basis of the theoretical and simulation results that we
presented in this paper for the dynamics of the bulk hard sphere
fluid, we conclude that the dynamical test particle limit, combined
with DDFT, provides a reliable method for calculating the van Hove
(and other related) dynamical pair correlation functions at low and at
intermediate densities $\rho  \sigma^3 \lesssim 0.6$. In the previous
publication \cite{Archer2007dynamics} we have shown that the theory
may be applied in a fairly straightforward manner also to
inhomogeneous situations, hence we conclude that the dynamic test
particle theory may indeed be used to calculate the van Hove function
for fluids at interfaces and under confinement. Furthermore we have
shown that, quite surprisingly, at low and intermediate densities the
very simple Vineyard approximation (reviewed in
Sec.\ \ref{sec:vineyard_appr}) is actually quantitatively fairly
good. This approximation only requires as input the radial
distribution function $g(r)$ and the diffusivity $D_l$ and therefore
provides a very useful quick approach for obtaining an approximation
for the van Hove function for colloidal fluids.

The possible conclusions about the performance and even about the
qualitative status of the predictions of the theory at higher
densities are far more intricate though. In this regime, the theory in
its current form is clearly quantitatively unreliable -- compare for
example, the results from BD simulations in Fig.\ \ref{fig:sim}(d) to
those from the dynamic test particle theory shown in
Fig.\ \ref{fig:ddft}(d) for $\rho  \sigma^3 =0.8$, where the theory
predicts that the system jams, whereas in simulations the system is an ergodic fluid. Moreover, at even higher
densities, the monodisperse hard sphere system crystallises in
simulations rather than undergoing a glass transition; polydispersity
would be required to suppress freezing \cite{AF2001nature}. We believe that the
behaviour of the theory at higher densities can primarily be ascribed
to our choice of approximation for the free energy functional -- see
Sec.\ \ref{sec:RY_func} for the reasons for using the RY functional
\eqref{eq:Fex_def} in the present study. We believe that if we had
used a more accurate functional, such as Rosenfeld's fundamental
measure theory \cite{rosenfeld89,tarazona08review,roth10review}, quantitatively
more accurate results could be obtained at these higher densities.
Nevertheless, the theory yields a clear prediction that there is a
dynamic (glass) transition, whereby the tagged (self) particle becomes
trapped in the cage formed by the surrounding particles. We can offer
three possible interpretations of this result.

Firstly, one may conclude that this glass transition stems simply from
the use of the approximate RY functional and since the theory predicts
the glass transition to be at a density value that is well below where
the glass transition is believed to occur, our results at higher
densities should be disregarded.

An alternative conclusion that one may draw is that the theory is
correctly describing (some of) the physics of the glass transition,
but that the predictions are only qualitative in nature and occur in
reality at higher densities and in polydisperse systems. Support for this point of view comes in
particular from results such as those in Fig.\ \ref{fig:intermediate},
where we display results for $\phi(k\sigma=2\pi,t)$. For $\rho 
\sigma^3=0.7$, which is near to where the theory predicts the glass
transition to be, we find a plateau in $\phi(k,t)$ and the clear
presence of two-stage relaxation, which indeed has been observed in
hard sphere colloidal suspensions \cite{vanmegen1993gtc}. Hence our
results are (qualitatively) similar to those from MCT \cite{Gotze1989, vanmegen1993gtc}.
Further support for the above interpretation comes from the results in
Figs.\ \ref{fig:energy_landscape} and
\ref{fig:vineyard_energy_landscape} for the behaviour of the free
energy landscape that underlies the DDFT. The appearance of a minimum
in the free energy as a function of displacement, corresponding to a
particle becoming trapped in the cage formed by its neighbours, is one
central prediction of the theory by Schweizer and Saltzmann
\cite{saltzman2003tcg,schweizer2005dmt}, who combined elements of MCT,
DFT, and activated rate theory, in order to describe localization and
transport in glassy fluids. Furthermore, given some of the work in the
literature based on DFT to study the glass transition, our prediction
that particles become localized should not come as a surprise: Wolynes
and coworkers \cite{stoessel1984lea, singh1985hsg,xia2001mth}
developed a successful model of hard sphere vitrification, which is
similar to the DFT treatment of crystallization
\cite{ramakrishnan1979fpo, hansen2006tsl}. Using a random
close-packed, non-periodic lattice they found a fluid-glass transition
where the fluid ``crystallizes'' onto this lattice. The success of
this method, along with its ability to model the freezing transition
(onto a regular lattice), has provoked a number of further
developments
\cite{kim2003gth,kaur2001hsl,kaur2002msm,baus1986hsg,lowen1990ech}. Other
approaches \cite{lust1993nhh} have investigated dense Brownian systems
through modelling via certain stochastic differential equations and
found that the system exhibits glassy behaviour. Thus, overall our
results seem to be qualitatively consistent with other DFT
based theories and with MCT for the glass transition. Nevertheless, the density
where the glass transition occurs, as predicted by the theory in its
present form, is far too low. Furthermore, it could be the case that the similarity between our results and those from the MCT are somehow a mathematical (rather than physical) coincidence, since an essential feature of MCT is the presence of memory in the dynamical equations. This important feature is absent from the present DDFT.

The third possible conclusion that one may draw concerns the question
whether the minimum in the free energy and the localization of the
tagged self particle are merely a signature of freezing in the
theory. The RY functional is well-known to predict the freezing
transition to occur at a density below that where it occurs in
reality. It could simply be the case that this functional overly
favours freezing, so that when it is applied in the way we use it
here, where we constrain all density profiles, $\rho_s(r)$ and
$\rho_d(r)$, to be spherically symmetric, a signature of freezing
shows up as the tagged self particle becoming localized.

Some merit can be found in all of the arguments outlined above and we
find ourselves unable to judge which one(s) are correct. Indeed
further work is required to provide a clear assessment of these
issues. In particular, the dynamical test particle theory should be
implemented with a more sophisticated approximation for the free
energy functional than we have used here.

As we have shown, our approach is based on integrating the Smoluchowski equation \eqref{eq:smoluchowski} over all
except one of the position coordinates, in order to derive an equation for the one-body density distribution. An alternative approach is to integrate over all but two of the position coordinates, in order to obtain \eqref{eq:intrho}
an equation for the two-body distribution function
$\rho^{(2)}(\rr_1,\rr_2,t)=N(N-1)\int  \int \dr_3 \, ... \int \dr_N P(\rr^N,t)$.
The resulting dynamical equation depends on the three body distribution function
$\rho^{(3)}(\rr_1,\rr_2,\rr_3,t)$. On making a suitable closure approximation,
this provides a different starting point for studying the pair correlations in a
colloidal fluid -- see e.g.\ Refs.\ \onlinecite{Brady1,Brady2} and
references therein, which also consider the effect of the
hydrodynamic interactions between the colloids. Developing the theory
for the dynamical pair correlation functions in this way is very natural. However, we believe that the strength of our method, where we use the dynamical test particle approach allowing us to work at the one-body level, is that we are able to use DFT to close our equations and therefore we are able to describe the fluid spatial correlations very accurately.

Finally, we mention other possible directions for developing the
theory in the future. One important aspect in the dynamics of
colloidal dispersions, that we have entirely neglected here, are the
hydrodynamic interactions between the particles. Rex and L\"owen
\cite{RL2008prl,RL2009epe} have shown how to include the hydrodynamic
interactions in a DDFT treatment and so it would be worthwhile to use
their DDFT formulation together with the present dynamical test
particle limit, in order to calculate the van Hove function under the
influence of hydrodynamic interactions.

A further aspect of our work,
that offers possible extensions of the theory, concerns the question
how to model the diffusivity of the tagged particle in a better way:
In the dynamical test particle calculation one could replace the
(constant) diffusion coefficient in Eq.\ \eqref{eq:DDFT_TP} with a
diffusion coefficient that depends on time; i.e.\ to replace $D \to
D(t)$. In doing this one could ensure that $D(t)$ takes the correct
values at both short and long times. However, doing this still does not
treat memory effects in the dynamics. As MCT demonstrates, memory
effects are key for a system to exhibit the ideal glass transition scenario\cite{gotzelmann1997dpa,vanmegen1993gtc,Nageleetal}.
Thus, we believe that including memory into our theory would be a crucial
step in future work. This could possibly be done
along the lines of the interesting work of Medina-Noyola and coworkers
\cite{YRMN2000pre,YRMN2001pre, MN2003pre, CRMN2006physicaA,MN2009jpcm}.
To include memory in our theory one could replace Eq.\ (29)
with\cite{MN2009jpcm,MedinaNoyola_privcom,Koideetal}:
\begin{align}
\frac{\partial \rho_i(\rr,t)}{\partial t} &= &\nabla \cdot
\int_0^t dt' \int \dr' \Gamma(\rr-\rr',t-t') \notag \\ & & \times 
 \left[\rho_i(\rr',t') \nabla \frac{\delta F[\{\rho_i\}]}{\delta
\rho_i(\rr',t')}\right],
\label{eq:MT_DDFT2}
\end{align}
where the mobility coefficient $\Gamma$ has been replaced by one that is
non-local in time and space. However, this would result in a considerable increase in computational complexity as within DFT the correlations in space are already treated in a complex manner and these would need to be coupled to the correlations in time. Whether such non-locality helps to cure some of the deficiencies of our approach is an open question. 

\appendix
\section{Exact Results}
\label{app:A}

We consider a fluid of $N$ particles with positions $\rr_i$, momenta $\pp_i$ and mass $m$ in the presence of an arbitrary external field that acts {\it only} on particle $i=1$, $u_1(\rr)=\lambda\rr_1^2$. Assuming that we are in the classical limit, the Hamiltonian is given by $H_N = K + V + U$ where the contributions are due to the (classical) kinetic energy, the total inter-particle potential (not necessarily pairwise
additive), and the external potential, respectively;
\begin{eqnarray}
K &=& \sum_{i=1}^N\frac{\pp_i^2}{2m} \\
V &=& v(\rr_1,..,\rr_N) \\
U &=& \lambda\rr_1^2. \label{eq:our_extpot}
\end{eqnarray}
The {\it canonical partition function}, $Q_N(V,T)$, is given by
\begin{equation}
Q_N(V,T)=\frac{h^{-3N}}{(N-1)!}\int\int\drr^N\dd \pp^N\exp[-\beta H_N(\rr^N,\pp^N)]
\label{eq:app_cpf}
\end{equation}
where $h$ is Planck's constant, and the $(N-1)!$ factor results from the fact that besides particle $i=1$, the remaining particles are indistinguishable. The integrations over momenta in Eq.\ \eqref{eq:app_cpf} can be carried out explicitly, leaving a {\it configuration integral} over positional degrees of freedom:
\begin{equation}
Z_N = \int\drr_1...\drr_N \exp(-\beta (V + U)).
\label{eq:Z_N}
\end{equation}
Note that for Brownian particles $Z_N$ is also the quantity that characterises the structure of the fluid.
Substituting our external potential \eqref{eq:our_extpot} into \eqref{eq:Z_N} we obtain
\begin{eqnarray*}
Z_N &=& \int\drr_1\exp(-\beta\lambda\rr_1^2)
    \int\drr_{2..N} \exp(-\beta V(\rr^N)) \\
 &=& \int\drr_1\exp(-\beta\lambda\rr_1^2)
    \int\drr'_{2..N} \exp(-\beta V(\rr'_{2..N})),
\end{eqnarray*}
where in the second step we have made the substitution $\rr'_i=\rr_i-\rr_1$, for $i=2..N$, so that we can do the integrations over the positions $\rr'_2...\rr'_N$. This gives,
\begin{eqnarray*}
Z_N &= \int\drr_1\exp(-\beta\lambda\rr_1^2) Z'_N
 &= (\pi/\beta\lambda)^{3/2} Z'_N,
\end{eqnarray*}
where $Z'_N$ is the configuration integral for $N$ particles where one particle is located at the origin. The Helmholtz free energy is then given by, $F=-\beta^{-1}\ln(Q_N(V,T))=-\beta^{-1}\ln(Q_N^{\rm id}\frac{Z_N(V,T)}{V_N})$ where $V_N$ is the volume occupied by the particles, which yields
\begin{equation}
\beta F = F_{\rm id} - \ln(Z'_N/V_N)-\frac{3}{2}\ln\left(\frac{\pi}{\beta\lambda}\right).
\label{eq:app_F}
\end{equation}
Therefore, the Helmholtz free energy only depends on the confining potential in a simple way.

One can also obtain the one body density profiles. In general, for a system of $N$ particles, the one body density profile, $\rho_N^{(1)}(\rr)$ can be obtained from~\cite{hansen2006tsl},
\begin{equation}
\rho_N^{(1)}(\rr)=\frac{N!}{Z_N(N-1)!}\int\drr^{(N-1)}\exp\left[-\beta (V(\rr^N) + \Phi(\rr^N))\right],
\end{equation}
where the $N!/(N-1)!$ factor accounts for the indistinguishability of the particles.

For the single particle subject to the external potential we get
\begin{eqnarray*}
\rho_1^{(1)}(\rr_1) &=& \frac{\exp(-\beta\lambda\rr_1^2)}{Z_N}\int\drr_{2..N}\exp(-\beta V(\rr^N)), \\
& =& \frac{\exp(-\beta\lambda\rr_1^2)}{(\pi/\beta\lambda)^{3/2}Z'_N}\int\drr'_{2..N}\exp(-\beta V(\rr'_{2..N})), \\
& =& \frac{\exp(-\beta\lambda\rr_1^2)}{(\pi/\beta\lambda)^{3/2}},
\end{eqnarray*}
which is a normalised Gaussian. It can be shown that since $\rho_s(r)$ is a Gaussian, then $w$ and $\lambda$ are simply related by
\begin{equation}
w = (2 \lambda /3)^{-1/2},
\end{equation}
and we can rewrite Eq.\ \eqref{eq:app_F} as 
\begin{equation}
F = F_{\rm id} - \ln(Z'_N)-\frac{3}{2}\ln\left(\frac{2\pi w^2}{3\beta}\right).
\label{eq:app_F2}
\end{equation}
We now seek the density profile of the remaining particles:
\begin{eqnarray}
\rho_2^{(1)}(\rr_2) & =&\frac{(N-1)!}{Z_N(N-2)!} \int\drr_1 \exp(-\beta\lambda\rr_1^2) \nonumber \\ && \quad \quad \quad\times	
 \int\drr_{3..N}\exp(-\beta V(\rr^N)), \nonumber\\
 & =&\frac{\int\drr_1 \exp(-\beta\lambda\rr_1^2)}{(\pi/\beta\lambda)^{3/2}} \label{eq:app_rho2} \\ && \times \frac{(N-1)}{Z'_N}	
 \int\drr_{3..N}\exp(-\beta V(\rr^N)) \nonumber .
\end{eqnarray}
To progress we make use of the formal relationship between $g(r)$ and the two body density profile, $\rho_N^{(2)}(\rr_1,\rr_2)$, which for a homogeneous fluid can be shown to be~\cite{hansen2006tsl}, 
\begin{eqnarray*}
g_N^{(2)}(\rr_1-\rr_2)&=&\frac{1}{\rho^2}\rho_N^{(2)}(\rr_1-\rr_2), \\
& =& \frac{1}{\rho^2}\frac{N!}{Z_N(N-2)!}\\
&&\quad \quad \quad \times
\int\drr_{3..N}\exp(-\beta V(\rr^N)), \\
& =& \frac{V}{\rho N}\frac{N(N-1)}{Z'_NV}\\
&&\quad \quad \quad \times
\int\drr_{3..N}\exp(-\beta V(\rr^N)),
\end{eqnarray*}
where we have made the substitutions, $\rho=N/V$ and $Z_N=Z'_NV$. Cancelling terms and rearranging we get,
\begin{equation}
\rho g_N^{(2)}(\rr_1-\rr_2)=\frac{(N-1)}{Z'_N}\int\drr_{3..N}\exp(-\beta V(\rr^N)).
\label{eq:app_gr}
\end{equation}
Substituting \eqref{eq:app_gr} into \eqref{eq:app_rho2} gives,
\begin{eqnarray*}
\rho_2^{(1)}(\rr_2) & =& \frac{\int\drr_1 \exp(-\beta\lambda\rr_1^2)}{(\pi/\beta\lambda)^{3/2}}\rho g_N^{(2)}(\rr_1-\rr_2) \\
& =&\frac{\rho}{(\pi/\beta\lambda)^{3/2}}\int\drr_1 \exp(-\beta\lambda\rr_1^2)g_N^{(2)}(\rr_1-\rr_2),
\end{eqnarray*}
which is the normalised Gaussian convolved with $\rho g(r)$.

\section*{Acknowledgements}

PH thanks the EPSRC for funding under grant EP/E065619/1 and AJA gratefully acknowledges financial support from RCUK. MS and AF thank DFG for support via SFB840/A3.


\begin{thebibliography}{78}
\bibitem[{foo({\natexlab{b}})}]{footnote2}
\bibinfo{note}{We have also calculated the external potentials corresponding to
  using the zero-dimensionality route for calculating $g(r)$ and found that the
  external potentials are largely unchanged. However, in the zero
  dimensionality route the magnitude of the repulsion in $u_d(r,t)$ is much
  smaller which results in more infilling in the core region.}
\end{thebibliography}

\end{document}